\documentclass[10pt, conference, letterpaper]{IEEEtran}
\IEEEoverridecommandlockouts
\usepackage{cite}
\usepackage{amsthm}
\usepackage{amsmath,amssymb,amsfonts}
\usepackage[noend]{algorithmic}
\usepackage{pifont}
\usepackage{dsfont}

\usepackage{graphicx}
\usepackage{float}

\usepackage{multirow}

\usepackage{textcomp}
\usepackage{xcolor}
\def\BibTeX{{\rm B\kern-.05em{\sc i\kern-.025em b}\kern-.08em
    T\kern-.1667em\lower.7ex\hbox{E}\kern-.125emX}}

\usepackage{float}
\usepackage{algorithm}
\usepackage{subfigure}
\usepackage{makecell}
\usepackage{wrapfig}
\usepackage{rotating}

\usepackage{xcolor}

\usepackage{tabularx}

\usepackage{fancyhdr}

\newtheorem{definition}{Definition}
\newtheorem{theorem}{Theorem}

\begin{document}
	\bstctlcite{IEEEexample:BSTcontrol}

\title{Multi-class Item Mining under Local Differential Privacy
}

\author{
\IEEEauthorblockN{Yulian Mao\textsuperscript{1,2}, Qingqing Ye\textsuperscript{2*}, Rong Du\textsuperscript{2}, Qi Wang\textsuperscript{1*}, Kai Huang\textsuperscript{3},  Haibo Hu\textsuperscript{2}}\\
\textsuperscript{1}{\small Department of Computer Science and  Engineering, Southern University of Science and Technology} \\
\textsuperscript{2}{\small Department of Electrical and Electronic Engineering, The Hong Kong Polytechnic University}\\
%\textsuperscript{3}{\small Research Institute of Trustworthy Autonomous Systems, Southern University of Science and Technology}\\
\textsuperscript{3}{\small School of Computer Science and Engineering, Macau University of Science and Technology}\\
{\{\small yulian.mao, roong.du\}@connect.polyu.hk, \{qqing.ye, haibo.hu\}@polyu.edu.hk,} {\small wangqi@sustech.edu.cn, kylehuangk@gmail.com}
}

\IEEEaftertitletext{\vspace{-1\baselineskip}}

\maketitle
\thispagestyle{fancy}
\fancyhead[C]{\textcolor{red}{This paper has been accepted by IEEE 41st Annual International Conference on Data Engineering (ICDE2025)}}

\begin{abstract}
%Item mining is a fundamental task in the context of local differential privacy (LDP), focusing on aggregating statistical information from users.
Item mining, a fundamental task for collecting statistical data from users, has raised increasing privacy concerns. To address these concerns, local differential privacy (LDP) was proposed as a privacy-preserving technique.  Existing LDP item mining mechanisms primarily concentrate on global statistics, i.e., those from the entire dataset. Nevertheless, they fall short of user-tailored tasks such as personalized recommendations, whereas classwise statistics can improve  task accuracy with fine-grained information. Meanwhile, the introduction of class labels brings new challenges. Label perturbation may result in invalid items for aggregation. To this end, we propose frameworks for multi-class item mining, along with two mechanisms: validity perturbation to reduce the impact of invalid data, and correlated perturbation to preserve the relationship between labels and items. We also apply these optimized methods to two multi-class item mining queries: frequency estimation and top-$k$ item mining. Through theoretical analysis and extensive experiments, we verify the effectiveness and superiority of these methods.
\end{abstract}

\begin{IEEEkeywords}
Local differential privacy, Multi-class item mining, Frequency estimation, Top-$k$ item mining
\end{IEEEkeywords}
\footnotetext{* Corresponding authors}
%\vspace{-1em}
%\footnotetext{* Dr. Qingqing Ye is the corresponding author.}

\section{Introduction}
With the rise of big data analytics, item mining has become essential for collecting valuable statistics to enhance model performance~\cite{amazon2017}. During the data collection process, users' private information may be at risk of exposure~\cite{Narayanan2008}. To address this privacy concern, differential privacy~\cite{dwork2006differential} has been widely employed to ensure rigorous privacy guarantees.
Its variant, namely local differential privacy (LDP), eliminates the requirement of a trusted third party for distributed data aggregation~\cite{kasiviswanathan2011can, ye2020local}, where local noise is added to users' data before publication. The advent of LDP has enabled rigorous privacy-preserving statistic estimation such as mean and frequency~\cite{erlingsson2014rappor, appleprivacy,  wang2019collecting}, based on which item mining becomes prevalent ~\cite{wang2017locally, ye2019, wang2018itemset, li2020estimating}. For instance, Google integrates RAPPOR into Chrome browser to collect web data~\cite{erlingsson2014rappor}, and Apple uses HCMS mechanism to gather emoji usage statistics from keyboard inputs~\cite{appleprivacy}.

Existing item mining mechanisms under LDP primarily focus on global statistics derived from the entire dataset. However, in many real-life applications in recommendation and machine learning systems, statistics from specific ``classes" can provide fine-grained information and thus enhance task accuracy. 
The following two examples show the potential applications of multi-class item mining.

%\begin{itemize}
%	\item \textbf{Shopping preference across user groups.} Study has shown that online shopping preference varies based on personal attributes such as gender and age~\cite{cole2008decision, zhou2014moderating, lian2014online, fang2016consumer}. Consequently, for personalized recommendations, item statistics ``grouped by" these attributes are more precise than global statistics derived from the entire dataset.
%	\item \textbf{Feature selection for different classes.} Many machine learning tasks rely on good feature selection to identify classwise characteristics~\cite{li2017feature}. For instance, decision tree branch selection based on information gain relies on the item frequency within each class~\cite{li2017feature}~\cite{quinlan1986induction}.
%\end{itemize}

\begin{itemize}
	\item \textbf{Shopping preference across user groups.} Studies have shown that personal attributes, such as gender and age, reflect the shopping preference~\cite{cole2008decision, zhou2014moderating, lian2014online, fang2016consumer}. Consequently, for recommendation systems, item statistics ``grouped by" these attributes are more precise than global statistics derived from the entire dataset.
	\item \textbf{Patients' data for disease diagnosis.} Machine learning plays a critical role in early-stage disease diagnosis~\cite{kononenko2001machine, ahsan2022machine}. To identify disease symptoms, statistics from medical test data need to be collected for model training. Consequently, the collected data should be associated with labels indicating whether the data belongs to a healthy individual or a patient.
\end{itemize}

Therefore, accurately estimating the statistics of multi-class items is crucial under LDP.
In this paper, we formulate such a problem, where each user holds a label-item pair that indicates the item belongs to the class label. Our goal is to mine classwise statistics from these label-item pairs under LDP. In particular, we aim to gather the item statistics within each class by aggregating label-item pairs from users using LDP mechanisms.
%[TODO: up to this line, there's no definition of multi-class statistics yet. I also suggest ``classwise statistics". ]
%Compared to existing item mining tasks that obtain statistics from the entire dataset, the key challenge in multi-class item mining lies in identifying the relationship between items and their corresponding labels.

An intuitive solution framework is to treat multi-class item mining as a series of item mining tasks across different classes by user partition~\cite{erlingsson2014rappor, wang2018itemset, ren2022ldp}.  Specifically, users are grouped according to the number of classes, with each group mining the items for a specific class.
However, this framework has a notable limitation: most users in each group may not possess the target label, leading to a significant proportion of invalid users. 
%For instance, if data is uniformly distributed across $c$ classes, about $\frac{c-1}{c}$ of users will be invalid, resulting in substantial noise that obscures the true statistics.

In essence, the key challenge in multi-class item mining lies in identifying the relationship between items and their corresponding labels. To preserve such a relationship, we propose two new frameworks to perturb the label-item pairs. The first framework perturbs the label-item pairs \textit{jointly} through expanding the perturbation domain to the Cartesian product of the item and label domains. Then each label-item pair is perturbed as a whole within this enlarged domain.
In contrast, the second framework perturbs labels and items \textit{separately}.
However, using existing perturbation mechanisms on these frameworks may still incur invalid data. For instance, the item becomes meaningless if the label is perturbed to other values.

To address this, we propose two perturbation mechanisms to process invalid data and efficiently preserve the relationship between items and labels. To mitigate the impact of invalid data, we introduce \textit{validity perturbation mechanism},  which reduces the noise injected by invalid data. Based on this, we propose \textit{correlated perturbation mechanism} to effectively preserve the relationship between labels and items by checking the perturbed labels before item perturbation. We then apply these optimized methods to multi-class item mining queries, including frequency estimation and top-$k$ item mining. %In addition to deriving unbiased frequency estimation mechanisms, we propose optimized approaches specifically tailored for top-$k$ item mining. 
Our contributions are as follows:
\begin{itemize}
	\item To the best of our knowledge, this is the first study to address item mining in the multi-class context. To preserve the relationship between labels and items, we propose two foundational frameworks for aggregating label-item statistics.
	\item To enhance the utility of the frameworks, we introduce two optimized mechanisms as the perturbation module: validity perturbation mechanism which reduces noise injected from invalid data, and correlated perturbation mechanism that maintains the label-item relationship during perturbation.
	\item Leveraging on the proposed frameworks and mechanisms, we explore two typical types of item mining queries. We derive unbiased frequency estimations and propose optimized methods specifically tailored for top-$k$ item mining. Theoretical analysis and extensive experiments on both real-world and synthetic datasets show the effectiveness of our approach.
\end{itemize}

The remainder of this paper is organized as follows. Section~\ref{statement} provides the preliminary on LDP and formulates the problem. Section~\ref{baseline} presents two frameworks, followed by the optimized mechanisms in Section~\ref{improvement}. Section~\ref{analysis} provides a theoretical comparison between our proposed mechanisms and existing approaches. Section~\ref{application} discusses two item mining queries using the proposed methods. Section~\ref{experiments} presents the experimental results. Section~\ref{relatedworks} reviews the related literature, and Section~\ref{conclusion} concludes this paper.
\section{Preliminaries and Problem Formulation}\label{statement}
In this section, we first introduce the fundamental concepts of local differential privacy and the mechanisms used for frequency estimation. Following this, we present a formal problem definition.

\subsection{Local Differential Privacy}
Differential privacy (DP) was introduced as a privacy measure technique with theoretical guarantee. The formal definition is as follows.
\begin{definition}[Differential Privacy~\cite{dwork2006differential}]
	Given any two neighboring datasets $D$ and $D'$ with one record difference, a randomized mechanism $\mathcal{A}$ satisfies $\epsilon$-differential privacy if for all possible outputs $S\subseteq Range(\mathcal{A})$, $	\mathrm{Pr}[\mathcal{A}(D)\in S]\le e^\epsilon\times \mathrm{Pr}[\mathcal{A}(D')\in S],$
	%	\begin{equation}\nonumber
		%		\mathrm{Pr}[\mathcal{A}(D)\in S]\le e^\epsilon\times \mathrm{Pr}[\mathcal{A}(D')\in S],
		%	\end{equation}
	where $\epsilon$ is the privacy budget.
\end{definition}
Differential privacy ensures that an adversary cannot distinguish between records from two neighboring datasets, effectively concealing individual records within the overall dataset. However, a trusted third party is required to aggregate user data under DP, which may not be feasible in many real-world scenarios. To this end, local differential privacy (LDP)~\cite{kasiviswanathan2011can, ye2020local} was introduced. In LDP, each user's data is perturbed locally before uploading to an untrusted third party, ensuring that an adversary cannot distinguish between any two individual reports. The formal definition of LDP is as follows:
\begin{definition}[Local Differential Privacy~\cite{kasiviswanathan2011can, ye2020local}]
	Given any two inputs $v$ and $v'$ from the domain, a randomized mechanism  $\mathcal{A}$ satisfies $\epsilon$-local differential privacy if for all possible outputs $V\subseteq Range(\mathcal{A})$, $\mathrm{Pr}[\mathcal{A}(v)\in V]\le e^\epsilon\times \mathrm{Pr}[\mathcal{A}(v')\in V],$
	%	\begin{equation}\nonumber
		%		\mathrm{Pr}[\mathcal{A}(v)\in V]\le e^\epsilon\times \mathrm{Pr}[\mathcal{A}(v')\in V],
		%	\end{equation}
	where $\epsilon$ is the privacy budget.
\end{definition}

\subsection{Statistical Estimations under Local Differential Privacy}
%Statistical estimations are fundamental applications in the context of local differential privacy, including count and frequency estimations.  Specifically, the count of a value $v$ in the dataset $D$ is given by
%\begin{equation}\nonumber
%	\text{count}(v)=\sum_{d_i \in D}\mathds{1}(v),
%\end{equation}
%where $\mathds{1}(v)$ is the indicator function defined as
%\begin{equation}\nonumber
%	\mathds{1}(v)=\begin{cases}
	%		1 & \text{if } d_i=v \\
	%		0 & \text{if } d_i\ne v.
	%	\end{cases}
%\end{equation}
%Similarly, the frequency of a value $v$ in the dataset $D$ can be expressed as
%\begin{equation}\nonumber
%	f_v=\frac{1}{\lvert D\rvert}\sum_{d_i \in D}\mathds{1}(v).
%\end{equation}

Statistical estimations are fundamental tasks in the context of LDP, including mean and frequency estimations~\cite{wang2017locally, wang2019collecting, li2020estimating}.  Specifically, the frequency of a value $v$ in a given dataset $D$ is denoted as $f(v)=\sum_{v_i \in D}\mathds{1}_{v_i}(v),$
%\begin{equation}\nonumber
%	f(v)=\sum_{v_i \in D}\mathds{1}_{v_i}(v),
%\end{equation}
where $\mathds{1}_{v_i}(v)$ is the indicator function, which returns $1$ if  $v_i=v$, and returns $0$ otherwise.
%\begin{equation}\nonumber
%	\mathds{1}_{v_i}(v)=\begin{cases}
	%			1 & \text{if } v_i=v \\
	%			0 & \text{if } v_i\ne v.
	%		\end{cases}
%\end{equation}

One of the state-of-the-art mechanisms for frequency estimation under LDP, known as Optimal Unary Encoding (OUE)~\cite{wang2017locally, du2023ldptrace, huang2024ldpguard, hu2024real}, consists of two steps, namely encoding and perturbation.
\begin{itemize}
	\item \textbf{Encoding.} Given an item $v$ and the item domain $\{1, 2, \cdots, d\}$ with size $d$, the item is encoded into a $d$-bit vector $B=[b_1, b_2, \cdots, b_d]$, where only $b_v=1$ and all other bits are zero.
	\item \textbf{Perturbation.} Given a privacy budget $\epsilon$, each bit $B[i]$ at position $i$ will be perturbed to $B'[i]$ as
	\begin{equation}\nonumber
		\mathrm{Pr}[B'[i]=1]=\begin{cases}
			p=\frac{1}{2} & \text{if } b_i=1 \\
			q=\frac{1}{e^\epsilon+1} & \text{if } b_i=0.
		\end{cases}
	\end{equation}
\end{itemize}

Another commonly used mechanism is Generalized Random Response (GRR)~\cite{wang2017locally}. Given an item $v\in \mathcal{I}$ with domain size $d$ and the privacy budget $\epsilon$,  $v$ is perturbed to $v'$ as
\begin{equation}\nonumber
	\mathrm{Pr}[\mathrm{GRR}(v)=v']=\begin{cases}
		p=\frac{e^\epsilon}{e^\epsilon+d-1} & v'=v \\
		q=\frac{1}{e^\epsilon+d-1} & v'\in \{\mathcal{I}\setminus v\}.
	\end{cases}
\end{equation}
Wang {\it et al}.~\cite{wang2017locally} proposed an adaptive mechanism that selects either OUE or GRR based on item domain size $d$ to minimize variance: if $d<3e^\epsilon+2$, GRR is chosen; otherwise, OUE is chosen.

\subsection{Problem Definition}
Given a dataset $D$ consisting of $N$ users $\mathcal{U} = \{u_1, u_2, \dots, u_N\}$, $c$ classes $\mathcal{C} = \{C_1, C_2, \dots, C_c\}$, and $d$ items $\mathcal{I} = \{I_1, I_2, \dots, I_d\}$, each user $u_i$ possesses a label-item pair $(C, I)$ where $C \in \mathcal{C}$ and $I \in \mathcal{I}$.
Note that the item domain of each class is initially unknown. %, so each class’s item domain is assumed to encompass the entire set of items,  $\mathcal{I}$.
Given the dataset $D$, our goal is to perform two specific item mining tasks in a multi-class setting: frequency estimation and top-$k$ item mining.

\begin{definition}[Multi-class Frequency Estimation]
	Given a dataset $D$ composed of label-item pairs from $N$ users, i.e.,  $D=\{(C_{u_1}, I_{u_1}), (C_{u_2}, I_{u_2}), \cdots, (C_{u_N}, I_{u_N})\}$, where each user $u_i$ holds a label-item pair $(C_{u_i}, I_{u_i})$. The frequency of an item $I\in \mathcal{I}$ within a class $C\in \mathcal{C}$ is denoted as $f(C, I)=\sum_{u_i\in \mathcal{U}} \mathds{1}_{u_i}(C, I),$
	%	\begin{equation}\nonumber
		%		f(C, I)=\sum_{u_i\in \mathcal{U}} \mathds{1}_{u_i}(C, I),
		%	\end{equation}
	where $\mathds{1}_{u_i}(C, I)$ is an indicator function defined by
	\begin{equation}\nonumber
		\mathds{1}_{u_i}(C, I)=\begin{cases}
			1, & \text{if } C_{u_i}=C\land I_{u_i}=I,  \\
			0, & \text{if }  C_{u_i}\ne C \lor I_{u_i}\ne I.
		\end{cases}
	\end{equation}
	The estimated frequency of a label-item pair $(C, I)$ from a mechanism $\mathcal{A}$ is said to be unbiased if $\mathbb{E}[\mathcal{A}(C, I)]=f(C, I)$,
	%	\begin{equation}\nonumber
		%		\mathbb{E}[\mathcal{A}(C, I)]=f(C, I),
		%	\end{equation}
	where $\mathbb{E}(\cdot)$ denotes the expectation.
\end{definition}

\begin{definition}[Multi-class Top-$k$ Item Mining]
	Given a dataset $D$ from $N$ users holding label-item pairs, i.e.,  $D=\{(C_{u_1}, I_{u_1}), (C_{u_2}, I_{u_2}), \cdots, (C_{u_N}, I_{u_N})\}$, an item $I\in \mathcal{I}$ is a top-$k$ item within a class $C\in \mathcal{C}$ if it is among the $k$ most frequent items within that class.
\end{definition}

\subsection{HEC: A Strawman Solution}
In many item mining tasks, to avoid privacy budget allocation, user partition is commonly used, where users are divided into groups, with each group focusing on mining a particular item or set of items~\cite{erlingsson2014rappor, wang2018itemset, ren2022ldp}. In our multi-class item mining problem, a straightforward approach is to divide users by class and aggregate item information within each class independently.

\textbf{Handling each class independently (HEC).} The users are divided into $c$ groups corresponding to the number of classes. Within each group, item statistics of the assigned class are collected using an LDP mechanism with privacy budget $\epsilon$.
%\sout{The perturbation domain for items is defined as $\mathcal{I}'=\{I_1, I_2, \dots, I_d, \perp\}$, where $\perp$ denotes items whose labels do not match the assigned class.}
If a user's label does not match the assigned class, her item is considered invalid for that class. To comply with the LDP guarantee, she needs to randomly select an item from the item domain to ensure deniability.

Although HEC enables multi-class item collection, its effectiveness is limited because only users with items in the assigned class can contribute to item mining. This is especially true when the domain size $|\mathcal{I}'|=d$ is large, leading to a substantial amount of invalid data.

%its inefficiency arises from substantial wasted users, whose labels do not match the assigned class. In essence, HEC disregards the relationship between items and labels, resulting in a significant amount of invalid data.

\section{Frameworks for Multi-class Item Mining}\label{baseline}
In multi-class item mining, it is crucial to consider class information. With the introduction of class labels, it becomes essential to maintain the relationship between a label and the corresponding item in a label-item pair. In this section, we present two frameworks specifically designed to preserve this relationship.

\subsection{Overview}
\begin{figure}[!htb]
	\centering
	\includegraphics[width=0.45\textwidth]{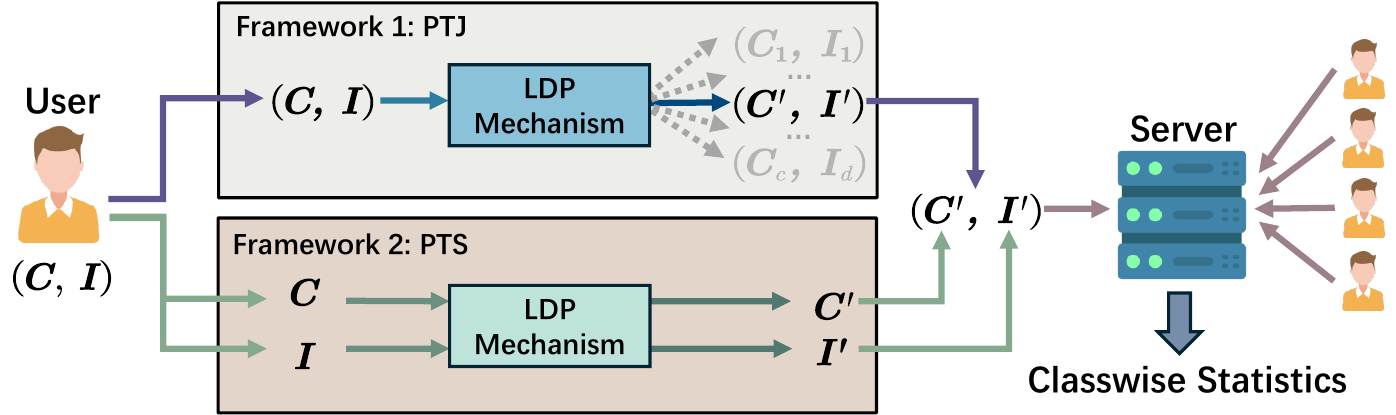}
	\caption{The overview illustrates two frameworks for multi-class item mining. The first framework, referred to as PTJ, treats a label-item pair as a whole and perturbs it to another pair. The second framework, PTS, perturbs each element separately.}
	\label{overview}
	\vspace{-1.5em}
\end{figure}

We propose two foundational frameworks designed to maintain the intrinsic relationship between labels and items. The first framework jointly perturbs the label and item (PTJ), maintaining their inherent connection. The second framework perturbs the label and item separately (PTS), followed by consolidated aggregation. Fig.~\ref{overview} illustrates both frameworks. Initially, a label-item pair is fed into one of the frameworks—either PTJ or PTS. Perturbation is then executed using a mechanism including existing LDP mechanisms, validity perturbation mechanism, and correlated perturbation mechanism. After perturbation, the perturbed pair is transmitted to the server for aggregation. Once all user data is aggregated, the server compiles the classwise statistics.

\subsection{Details of Frameworks}
As previously discussed, the HEC framework neglects the class information, leading to a significant amount of invalid users due to mismatched labels. To this end, an alternative approach treats the label-item pair as a cohesive unit, and then  jointly perturbs the entire pair.

\textbf{Perturbing the pair jointly (PTJ).} Given the label domain $\mathcal{C} = \{C_1, C_2, \dots, C_c\}$, and the item domain $\mathcal{I} = \{I_1, I_2, \dots, I_d\}$, the perturbation domain is defined as the Cartesian product $\mathcal{P}=\{(C_1, I_1), (C_1, I_2), \cdots\}$, with size $\lvert \mathcal{P} \rvert=c\times d$. Each user, holding a label-item pair, perturbs her pair using an LDP mechanism with privacy budget $\epsilon$ over the perturbation domain $\mathcal{P}$.
After aggregating all the perturbed pairs, the item information within each class can be inferred.

When either the label domain or the item domain is large, the size of the perturbation domain increases significantly, leading to high communication or computation cost~\cite{wang2017locally}. To mitigate this, an alternative approach is to perturb the label and item separately.

\textbf{Perturbing the pair separately (PTS).} For a user $u_i$ holding a label-item tuple $(C, I)$, the label is first perturbed via an LDP mechanism with part of the privacy budget $\epsilon_1$. Similarly, the item is perturbed using an LDP mechanism with the remaining privacy budget $\epsilon_2$.

Within these frameworks, existing perturbation mechanisms still suffer from invalid data in item mining tasks. For instance, the class label may be perturbed to other classes, resulting in an invalid item for that class. To enhance the utility of multi-class item mining, it is imperative to optimize perturbation mechanisms to process invalid information and account for the label-item relationship throughout the perturbation process.

%However, this approach also introduces significant noise. If the label is perturbed to other values, the item becomes irrelevant to the perturbed label. As a result, including such items in the aggregation adds noise. To improve the utility of multi-class item mining, it is crucial to account for invalid information and the inherent relationship between labels and items during the perturbation process.
%\vspace{-0.2em}
\section{Optimized Perturbations }\label{improvement}
In this section, we introduce the validity perturbation to process data including invalid items. Building on the validity perturbation, we propose the correlated perturbation to further preserve the label-item relationship during the perturbation process. In the interest of space, the proofs in
this section are included in our technical report~\cite{techr}.

\subsection{Validity Perturbation Mechanism}
In this subsection, we propose the validity perturbation mechanism based on the Unary Encoding (UE)~\cite{wang2017locally} mechanism. To process invalid items, an intuitive approach is to randomly select a valid item to replace the invalid one for perturbation and aggregation~\cite{wang2018privtrie, wang2021heavy}.
However, such random noise can distort the aggregated statistics of valid items, resulting in utility degradation. To address this, we should exclude those invalid items in the aggregation process. Specifically, we first privately publicize item validity, and then omit the invalid items. To avoid consuming an extra privacy budget for item validity, we integrate a validity flag into the UE mechanism to process invalid items. Both the validity flag and items are perturbed simultaneously.
%Avoiding consuming extra privacy budget, the perturbations on validity flag and item should occur simultaneously. Essentially, the two elements are bound together yet perturbed individually.
%Note that the hashing based mechanisms such as optimal local hashing (OLH)~\cite{wang2017locally} can not be adopted since we aim to omit the invalid items during aggregation.
The validity perturbation mechanism consists of encoding and perturbation.

\begin{figure}[!htb]
	\vspace{-1.5em}
	\centering
	\includegraphics[width=0.35\textwidth]{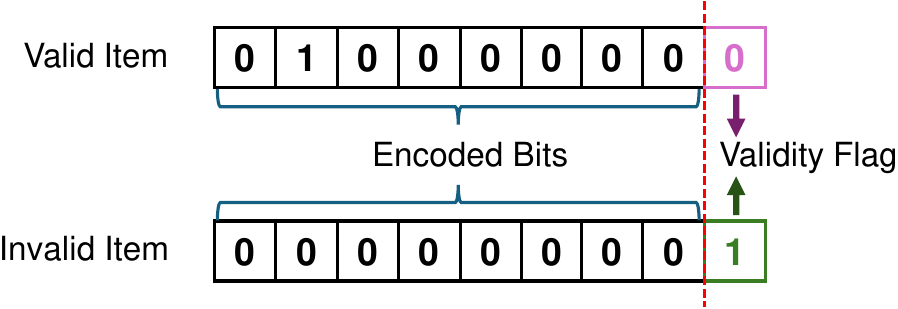}
	
	\vspace{-1.2em}
	\caption{An illustration for encoding scheme. For a valid item, the encoded bits from UE are padded with a validity flag ``0". Conversely, for an invalid item, all encoded bits are set to ``0", and the validity flag is set to ``1". }
	\label{validity_perturbation}
	\vspace{-0.5em}
\end{figure}

\textbf{Encoding.} As shown in Fig.~\ref{validity_perturbation}, given an item $v$ with domain size $d$, if the item is valid, $\mathrm{Encode}(v)=[0, \cdots, 0, 1, 0, \cdots, 0]$ is a $(d+1)$-length binary vector with the $v$-th position set to ``1". Conversely, if the item is invalid, $\mathrm{Encode}(v)=[0, \cdots, 0, 0, 0, \cdots, 1]$ is a $(d+1)$-length binary vector with the last position set to ``1". 

\textbf{Perturbing.} Given the encoded binary vector $B=\mathrm{Encode}(v)$, the output $B'$ from the perturbation $\mathrm{Perturb}(B)$ follows the same process as the UE mechanism~\cite{wang2017locally}. For any position $i$ in $B$, 
\vspace{-1em}
\begin{equation}\label{prob}
	\mathrm{Pr}[B'[i]=1]=\left\{
	\begin{aligned}
		p, & \quad \text{if $B[i]=1$} \\
		q,& \quad \text{if $B[i]=0$}.
	\end{aligned}
	\right.
\end{equation}
\vspace{-1.5em}
\begin{theorem}(Privacy of Validity Perturbation Mechanism)
	The validity perturbation mechanism satisfies $\epsilon$-LDP for $	\epsilon=\ln \frac{p(1-q)}{(1-p)q}$~\cite{wang2017locally}.
	%	\begin{equation}\nonumber
		%			\epsilon=\ln \frac{p(1-q)}{(1-p)q}.
		%	\end{equation}
\end{theorem}
%\begin{proof}
%Given any two items with validity, there are only two bits difference after encoding. Therefore, the $\epsilon$-LDP is guaranteed~\cite{wang2017locally}.
%\end{proof}

In this paper, the perturbation probabilities in Eq.~(\ref{prob}) are set the same as the Optimized Unary Encoding (OUE) mechanism for a convenient comparison with other LDP mechanisms, i.e., $p=\frac{1}{2}$, $q=\frac{1}{e^\epsilon+1}$, where $\epsilon$ represents the privacy budget~\cite{wang2017locally}.
A utility analysis of the validity perturbation mechanism is presented in Section~\ref{analysis}. We use the noise injected by invalid users as the utility metric~\cite{cao2021data}. The theoretical results demonstrate the superiority of the validity perturbation mechanism.

%\section{Correlated Perturbation}
With validity perturbation, we can effectively manage invalid data and reduce its impact. Based on this, we introduce the correlated perturbation, which preserves the relationship between labels and items during the perturbation process.

\subsection{Correlated Perturbation Mechanism}
In this subsection, we present the correlated perturbation mechanism. In multi-class item mining, the label and item are paired, exhibiting their inherent correlation. When the label is perturbed, the item should be changed accordingly, or labeled as noise to avoid unexpected influence. Therefore, the labels and the items should be perturbed in a correlated manner. Specifically, we can perturb the label first, and then perturb the item according to the label's perturbation result. If the label differs from its original value, the item becomes invalid and must be excluded from the aggregation result. For the purpose of LDP guarantee, an extra privacy budget is required to learn the condition of the perturbed label --- whether the perturbed label matches its original value. To save privacy budget, we adopt the validity perturbation mechanism, using the label condition as the validity flag. Based on this, we propose the correlated perturbation mechanism as follows.

%	In this subsection, we present the correlated perturbation mechanism. For multi-class item mining, the label and item are paired. If the label is perturbed to other values, the item should be treated as noise with respect to the perturbed label. Therefore, the labels and the items should be perturbed separately but correlated. That is, the label is perturbed first, and then the item is perturbed based on the label's perturbation result. If the label differs from its original value, the item becomes invalid and must be excluded from the aggregation result. For the purpose of LDP guarantee, extra privacy budget is required to learn the the condition of the perturbed label --- whether the perturbed label matches its original value. To save privacy budget, we adopt validity perturbation mechanism, using the label condition as the validity flag. Based on this, we propose the correlated perturbation as follows.

Given a label-item pair from a user with the class label $C\in \mathcal{C}$ $( \mathcal{C}=\{C_1, C_2, \cdots, C_c\})$ and the item $I\in \mathcal{I}$ $(\mathcal{I}=\{I_1, I_2, \cdots, I_d\})$, the perturbation process consists of label perturbation and item perturbation phases, where the total privacy budget $\epsilon$ is divided into $\epsilon_1$ for label perturbation and $\epsilon_2$ for item perturbation, such that $\epsilon=\epsilon_1+\epsilon_2$. In this paper, we set $\epsilon_1=\epsilon_2=\frac{\epsilon}{2}$.

\textbf{Label Perturbation.} For a given class label $C$ and privacy budget $\epsilon_1$, the label is perturbed by $\Phi$ with privacy budget $\epsilon_1$ such that
\begin{equation}\label{cpl}
	\mathrm{Pr}[\Phi(C)=C']=\left\{
	\begin{aligned}
		p_1 & \quad \text{if $C'=C$} \\
		q_1& \quad \text{if $C'\in \mathcal{C}\backslash\{C\}$},
	\end{aligned}
	\right.
\end{equation}
where $\Phi$ is an LDP mechanism, such as the Generalized Random Response (GRR) mechanism, and the ratio is $p_1/q_1\le\epsilon_1$ for LDP guarantee.

\textbf{Item Perturbation.} If the perturbed label differs from its original value, the item is marked as invalid; otherwise, the item is valid and the validity perturbation mechanism is applied to perturb the item with the privacy budget $\epsilon_2$.
For a given item $I$ from the item domain $\mathcal{I}$ with size $d$, and privacy budget $\epsilon_2$, the item will be first encoded to a bit vector $B$ with length $d+1$ according to the validity. Any position $j$ of $B$ after encoding is denoted as
\vspace{-0.5em}
\begin{equation}\nonumber
	B[j]=\left\{
	\begin{aligned}
		1 & \quad \text{if $(j=I\land C'=C)\lor (j=d+1\land C'\ne C)$} \\
		%		0 & \quad \text{if $(j\ne I\land C'= C)\lor (j\ne d+1\land C'\ne C)$}.
		0 & \quad \text{Otherwise}.
	\end{aligned}
	\right.
\end{equation}
Given the encoded bit vector $B$, any bit $B[j]\in B$ will be flipped via the OUE perturbation mechanism $\Psi$ with privacy budget $\epsilon_2$~\cite{wang2017locally},
\vspace{-1em}
\begin{equation}\label{cpi}
	\mathrm{Pr}[\Psi(B[j])=1]=\left\{
	\begin{aligned}
		p_2 & \quad \text{if $B[j]=1$} \\
		q_2& \quad \text{if $B[j]=0$},
	\end{aligned}
	\right.
\end{equation}

where $p_2=\frac{1}{2}$ and $q_2=\frac{1}{e^{\epsilon_2}+1}$.

\begin{theorem}(Privacy of Correlated Perturbation Mechanism)
	The correlated perturbation mechanism satisfies $\epsilon$-LDP. 
\end{theorem}
%\begin{proof}
%	The correlated perturbation mechanism consists of label perturbation and item perturbation, with privacy budgets $\epsilon_1$ and $\epsilon_2$, respectively. According to the principle of sequential composition theorem~\cite{li2017differential}, the correlated perturbation mechanism satisfies $(\epsilon_1+\epsilon_2)$-LDP, which is equivalent to $\epsilon$-LDP.
%\end{proof}

In certain item mining tasks, an unbiased frequency estimation is required~\cite{wang2017locally}. In the following, we present the calibration on correlated perturbation mechanism to obtain the unbiased result. 
Given the collected count $\tilde{f}(C, I)$ of a label-item pair $(C, I)$ aggregated from $N$ users, the calibrated result is 
\begin{equation}\label{unbiased_estimation}
	\begin{aligned}
		\hat{f}(C, I) =& \frac{\tilde{f}(C, I)-Nq_1q_2(1-p_2)}{p_1(1-q_2)(p_2-q_2)}\\
		&-\frac{\hat{n}q_2[p_1(1-q_2)-q_1(1-p_2)]}{p_1(1-q_2)(p_2-q_2)},
	\end{aligned}
\end{equation}
where $p_1$, $q_1$, $p_2$, and $q_2$ are the perturbation probabilities in Eqs.~(\ref{cpl}) and~(\ref{cpi}).  Additionally, $\hat{n}$, representing the unbiased frequency estimation of users with label $C$, is derived by  $	\hat{n}=\frac{\tilde{n}-Nq_1}{p_1-q_1},$
%		\begin{equation}\nonumber
	%				\hat{n}=\frac{\tilde{n}-Nq_1}{p_1-q_1},
	%			\end{equation}
where $\tilde{n}$ is the collected count of users with label $C$.

\begin{theorem}
The calibrated frequency $\hat{f}(C, I)$ is unbiased.
\end{theorem}

%	\begin{theorem}
%		The calibrated frequency $\hat{f}(C, I)$ is unbiased.
%	\end{theorem}

%	\begin{proof}
%		The collected count $\tilde{f}(C, I)$ of a label-item pair can be expressed as:
%		\begin{equation}\label{feeq}
	%			\begin{aligned}
		%				\tilde{f}&(C, I)=\underbrace{f(C, I)\cdot p_1\cdot (1-q_2)\cdot p_2}_{\text{\ding{192}}} \\
		%				&+ \underbrace{(n-f(C, I))\cdot p_1\cdot (1-q_2)\cdot q_2}_{\text{\ding{193}}}\\
		%%		\end{aligned}
	%%	\end{equation}
%%	\begin{equation}\label{feeq}
	%%		\begin{aligned}
		%				&+\underbrace{(\sum_{C_i\in \mathcal{C} }f(C_i, I)-f(C, I))\cdot q_1\cdot (1-p_2)\cdot q_2}_{\text{\ding{194}}}\\
		%				&+\underbrace{[N-n-(\sum_{C_i\in \mathcal{C} }f(C_i, I)-f(C, I))]\cdot q_1\cdot (1-p_2)\cdot q_2}_{\text{\ding{195}}},
		%			\end{aligned}
	%		\end{equation}
%		where $f(C, I)$ represents true frequency, $n$ denotes the number of users with class label $C$, and the term $\sum_{C_i\in \mathcal{C}}f(C_i, I)$ represents the total number of users with the item $I$.
%		Part \ding{192} refers to users who posses the pair $(C, I)$ and retain the pair after perturbation.
%		Part \ding{193} accounts for users with the same label but different items, who perturb their pairs to $(C, I)$.
%		Part \ding{194} represents users with the same item but different labels, who perturb their pairs to the target pair $(C, I)$.
%		Lastly, part \ding{195} covers users with different items and different labels who perturb their pairs to $(C, I)$. The unbiased frequency estimation $\hat{f}(C, I)$ can be derived from Eq.~(\ref{feeq}).
%	\end{proof}
\vspace{-1.2em}
\section{Utility Analysis}\label{analysis}
\vspace{-0.5em}
In this section, we theoretically analyze the utility of the proposed mechanisms, validity perturbation and correlated perturbation, to demonstrate their superiority. In the interest of space, the proofs in
this section are included in our technical report~\cite{techr}.

\subsection{Utility Analysis for Validity Perturbation Mechanism}
As aforementioned, the validity perturbation mechanism is to mitigate the impact of invalid data. We use the noise injected by invalid data as the utility metric~\cite{cao2021data}. 
The uploaded item from an invalid user can be regarded as a random injection. For instance, the item domain is narrowed down during the mining process, and infrequent items may be pruned. An invalid user is the one who possesses an infrequent item that has been pruned. To comply with LDP, the invalid user randomly selects a valid candidate for deniability~\cite{wang2018privtrie, wang2021heavy}. The injected noise can be derived as follows.
\begin{theorem}
	Given the valid item domain size $d$ and the number of invalid users $m$, the noise injected into a valid item from an LDP mechanism is $\mathbb{E}_{noise}=mq+\frac{1}{d}m(p-q),$
	%	\begin{equation}\nonumber
		%		\mathbb{E}_{noise}=mq+\frac{1}{d}m(p-q),
		%	\end{equation}
	and the variance of the injected noise is  $\text{Var}_{noise}=mq(1-q)+\frac{m}{d}(p-q)(1-p-q),$
	%	\begin{equation}\nonumber
		%		\text{Var}_{noise}=mq(1-q)+\frac{m}{d}(p-q)(1-p-q),
		%	\end{equation}
	where $p$ and $q$ are the perturbation probabilities set by an LDP mechanism. 
\end{theorem}
%\begin{proof}
%	An invalid user first randomly selects a valid item and then applies the LDP mechanism. The expected noise is then given by
%	\begin{equation}\nonumber
	%		\begin{split}
		%			\mathbb{E}_{noise}&=m\cdot \frac{1}{d}\cdot p+m\cdot (1-\frac{1}{d})\cdot q\\
		%			&=mq+\frac{1}{d}m(p-q).
		%		\end{split}
	%	\end{equation}
%	Furthermore, the variance of the injected noise is:
%	\begin{equation}\nonumber
	%		\begin{split}
		%			\text{Var}_{noise} &=m\cdot \frac{1}{d}\cdot p(1-p)+m\cdot (1-\frac{1}{d})\cdot q (1-q)\\
		%			& = mq(1-q)+\frac{m}{d}(p-q)(1-p-q).
		%		\end{split}
	%	\end{equation}
%\end{proof}

Similarly, the injected noise of the validity perturbation can be expressed as follows.

\begin{theorem}
	Given the valid item domain size $d$ and $m$ invalid users, the noise injected on a valid item via the validity perturbation mechanism is $\mathbb{E}_{noise}=mq(1-p)$,
	%	\begin{equation}\nonumber
		%		\mathbb{E}_{noise}=mq,
		%	\end{equation}
	and the variance of the injected noise is $	\text{Var}_{noise}=mq(1-q)-mpq(1+pq-2q)$,
	%	\begin{equation}\nonumber
		%		\text{Var}_{noise}=mq(1-q),
		%	\end{equation}
	where $q$ is the probability in Eq. (\ref{prob}). 
\end{theorem}

Clearly, the validity perturbation mechanism reduces the injected noise on valid items from invalid users.  The utility of the validity perturbation mechanism is superior to that of the OUE mechanism as a comparison~\cite{wang2017locally}, and by extension, it outperforms the other LDP mechanisms in processing invalid data.

\subsection{Effectiveness of Validity Perturbation}
Since the validity perturbation mechanism affects the perturbation results on both valid and invalid users, its effectiveness still needs to be evaluated. We first derive the expectation and variance of the collected counts using LDP mechanisms, and then compare these with the corresponding values derived from the validity perturbation mechanism.

\begin{theorem}
	Given that $N_1$ users hold the target item, $N_2$ users hold the other items in the valid item domain with size $d$, and $m$ users hold the invalid items.
	For an LDP mechanism, the count expectation of the target item is 
	\begin{equation}\nonumber
		\begin{split}
			\mathbb{E}(\text{count}) &=N_1p+N_2q+\frac{m}{d}p+m(1-\frac{1}{d})q\\
			&=N_1p+N_2q+mq+\frac{m}{d}(p-q).
		\end{split}
	\end{equation} 
	Correspondingly, the variance is 
	\begin{equation}\nonumber
		\begin{split}
			\text{Var(count)}&=N_1(p-p^2)+N_2(q-q^2)+m(q-q^2)\\
			& \quad\quad +\frac{m}{d}(p-q)(1-p-q).\\
		\end{split}
	\end{equation}
\end{theorem}  

Similarly, the expectation and the variance of the validity perturbation mechanism can be derived as follows.
\begin{theorem}
	Suppose $N_1$ users hold the target item, $N_2$ users hold the other items in the valid item domain with the size $d$, and $m$ users hold the invalid items.
	For the validity perturbation mechanism, the count expectation of the target item is 
	\begin{equation}\nonumber
		\begin{split}
			\mathbb{E}(\text{count}) &=N_1p\cdot (1-q)+N_2q\cdot (1-q)+mq\cdot (1-p)\\
			&=(1-q)(N_1p+N_2q+mq-mq\frac{p-q}{1-q}).
		\end{split}
	\end{equation} 
	And the variance is 
	\begin{equation}\nonumber
		\begin{split}
			\text{Var(count)}&=N_1(p-p^2+2p^2q-pq-p^2q^2)\\
			&+N_2(q-2q^2+2q^3-q^4)\\
			&+m(q-q^2+2pq^2-pq-p^2q^2).
		\end{split}
	\end{equation}
\end{theorem} 

To compare the validity perturbation mechanism with LDP mechanisms with invalid data, we use the Optimized Unary Encoding (OUE) mechanism as a comparison~\cite{wang2017locally}. In terms of expectations, the validity perturbation mechanism is preferable because it introduces less noise. Although the expectation is scaled, the counts of all items are scaled consistently, preserving the rank orders.
Regarding variance, the difference between the validity perturbation mechanism and the OUE mechanism   can be expressed as follows:
\begin{equation}\nonumber
	\begin{split}
		&N_1(2p^2q-pq-p^2q^2)+N_2(2q^3-q^2-q^4)\\
		&+m(2pq^2-pq-p^2q^2)-\frac{m}{d}(p-q)(1-p-q)\\
		=& N_1pq(2p-1-pq)+N_2q^2(2q-1-q^2)\\
		& \quad\quad +mpq(2q-1-pq) -\frac{m}{d}(p-q)(1-p-q),
	\end{split}
\end{equation}
which is always smaller than 0. Namely, the validity perturbation mechanism is better than the OUE mechanism to process data with invalid ones, thereby it is also better than the other LDP mechanisms.

\subsection{Utility Analysis for Correlated Perturbation}
In this section, we analyze the utility of the correlated perturbation mechanism with unbiased calibration for frequency estimation. To demonstrate the superiority of the correlated perturbation mechanism, we compare its variance with other LDP mechanisms to perturb the label-item pair. We first derive the variance of the correlated perturbation mechanism as follows.
\begin{theorem}
	The variance of the estimated frequency $\hat{f}(C, I)$ in Eq.~(\ref{unbiased_estimation}) is 
	\begin{equation}\label{varcp}
		\begin{aligned}
			Var[\hat{f}&(C, I)] =  \frac{f(C, I) p_1 (1-q_2) p_2 \left[ 1 - p_1 (1-q_2) p_2 \right]}{ \left[ p_1 (1-q_2)(p_2 - q_2) \right]^2 } \\
			& + \frac{(n - f(C, I)) p_1 (1-q_2) q_2 \left[ 1 - p_1 (1-q_2) q_2 \right]}{\left[ p_1 (1-q_2)(p_2 - q_2) \right]^2} \\
			& + \frac{(N - n) q_1 (1-p_2) q_2 \left[ 1 - q_1 (1-p_2) q_2 \right]}{\left[ p_1 (1-q_2)(p_2 - q_2) \right]^2} \\
			& + \left[ \frac{q_2 \left[ p_1 (1-q_2) - q_1 (1-p_2) \right]}{p_1 (1-q_2)(p_2 - q_2)} \right]^2 \\
			& \times \frac{n \left[ p_1 (1-p_1) - q_1 (1-q_1) \right] + N q_1 (1-q_1)}{(p_1 - q_1)^2},
			% Var&[\hat{f}(C, I)] = f(C, I)\frac{p_1 (1-q_2) (p_2-q_2) \left[ 1 - p_1 (1-q_2) p_2 \right]}{\left[ p_1 (1-q_2)(p_2 - q_2) \right]^2}\\
			% -n\frac{}{}
		\end{aligned}
	\end{equation}
	where $p_1$, $q_1$, $p_2$ and $q_2$ denote the perturbation probabilities in Eqs.~(\ref{cpl}) and~(\ref{cpi}). Additionally, $n$ represents the true number of users with label $C$, and $N$ refers to the total number of users. 
\end{theorem}
%\begin{proof}
%	According to Eq.~(\ref{unbiased_estimation}), the variance of the estimated frequency $\hat{f}(C, I)$  can be denoted as
%	\begin{equation}\label{proofvcp}
	%		\begin{aligned}
		%			Var[\hat{f}(C, I)]=&\underbrace{\frac{Var[\tilde{f}(C, I)]}{\left[ p_1 (1-q_2)(p_2 - q_2) \right]^2}}_{\text{\ding{192}}}\\
		%			&+\underbrace{\left[ \frac{q_2 \left[ p_1 (1-q_2) - q_1 (1-p_2) \right]}{p_1 (1-q_2)(p_2 - q_2)} \right]^2 Var[\hat{n}]}_{\text{\ding{193}}}\\
		%			=&\frac{Var[\tilde{f}(C, I)]}{\left[ p_1 (1-q_2)(p_2 - q_2) \right]^2}\\
		%			&+\left[ \frac{q_2 \left[ p_1 (1-q_2) - q_1 (1-p_2) \right]}{p_1 (1-q_2)(p_2 - q_2)(p_1-q_1)} \right]^2 Var[\tilde{n}].\\
		%		\end{aligned}
	%	\end{equation}
%	Thus, Eq.~(\ref{varcp}) can be derived straightforwardly by substituting $Var[\tilde{f}(C, I)]$ and $Var[\tilde{n}]$  into the equation above.
%\end{proof}

\textbf{Variance analysis.} 
Our investigation focuses on two critical aspects: the correlation strengths between labels and items, and the class distribution. To quantify label-item correlations, we employ pointwise mutual information (PMI)~\cite{church1990word}, defined as $\rm{PMI}(C;I)=\log_2\frac{\textit{p}(C, I)}{\textit{p}(C)\textit{p}(I)}$, where $p(C, I)$ denotes the joint probability of the label-item pair $(C, I)$, while $p(C)$ and $p(I)$ represent marginal probabilities of labels and items, respectively. A larger PMI indicates a stronger correlation between the label and the item. When $p(C)$ and $p(I)$ are fixed, we have $\rm{PMI(C;I)}\propto f(C, I)$. Since the variance in Eq.~(\ref{varcp}) is proportional to $f(C, I)$, it follows that  $Var[\hat{f}(C, I)]\propto \rm{PMI}(C;I)$.  Note that the variance equation exhibits complex structure. We numerically estimate the variable coefficients, which depend on the dataset,    under varying epsilon values to analyze variance dynamics. The coefficients are in Table \ref{coe}. Crucially, $f(C, I)$ is always much smaller than class amount $n$ and data amount $N$, so that the coefficients of $f(C, I)$ in Table~\ref{coe} cannot offset orders of magnitude differences. This magnitude disparity explains why correlation variations are concealed in variance analysis. The impact of class amount $n$ on variance can be analyzed in a similar manner. With fixed label-item pair frequency $f(C, I)$ and data amount $N$, we establish $Var[\hat{f}(C, I)]\propto n$, revealing that class diversity expansion directly amplifies variance. These theoretical predictions are empirically validated in Section \ref{experiments}.
\begin{table}[h]
	%	\centering
	\caption{Coefficients of variables in $Var[\hat{f}(C, I)]$}
	\label{tab:complexity}
	\begin{tabular}{c c ccccccc }  % l=左对齐, c=居中, r=右对齐
		\hline
		$\epsilon$ & $0.5$ & $1$ & $1.5$ & $2$ & $2.5$ & $3$ & $3.5$ & $4$\\ \hline
		$f(C, I)$ & 87.4& 32.9 & 17.1 & 10.3 & 6.8 & 4.9& 3.7 & 2.9 \\ 
		$n$ & 213.8 & 58.9 & 22.8 & 10.5 & 5.4 & 3.0 & 1.8 & 1.1 \\ 
		$N$ & 441.8 & 53.3 & 12.0 & 3.6 & 1.3 & 0.5 & 0.2 & 0.1 \\ \hline 
	\end{tabular}
	\label{coe}
	\vspace{-2em}
\end{table}

Similarly, the unbiased frequency estimation of a label-item pair and its corresponding variance under LDP mechanisms can be derived as follows.

\begin{theorem}
	Given a dataset $D$ consisting of $N$ users holding label-item pairs, using an LDP mechanism for label perturbation  with probabilities $p_1$ and $q_1$, and an LDP mechanism for item perturbation with probabilities $p_2$ and $q_2$, the estimated frequency $\hat{f}(C, I)$ of   an item $I$ within a class $C$ is given by:
	\begin{equation}\label{feoue}
		\begin{aligned}
			\hat{f}(C, I)=&\frac{\tilde{f}(C, I)-\hat{n}q_2(p_1-q_1)}{(p_1-q_1)(p_2-q_2)}\\
			&-\frac{\sum_{C_i\in \mathcal{C} }\hat{f}(C_i, I)q_1(p_2-q_2)+Nq_1q_2}{(p_1-q_1)(p_2-q_2)},
		\end{aligned}
	\end{equation}
	where $\tilde{f}(C, I)$ represents the collected count of the target label-item pair $(C, I)$, and 
	\begin{equation}\nonumber
		\sum_{C_i\in \mathcal{C} }\hat{f}(C_i, I)=\frac{\sum_{C_i\in \mathcal{C} }\tilde{f}(C_i, I)-Nq_2}{p_2-q_2}
	\end{equation}
	is the unbiased frequency estimation of the item $I$, where $\sum_{C_i\in \mathcal{C} }\tilde{f}(C_i, I)$ is the collected count of the item $I$.
	Additionally, $\hat{n}$, the unbiased frequency estimation of users with label $C$, is given by: $\hat{n}=\frac{\tilde{n}-Nq_1}{p_1-q_1},$
	%	\begin{equation}\nonumber
		%		\hat{n}=\frac{\tilde{n}-Nq_1}{p_1-q_1},
		%	\end{equation}
	where $\tilde{n}$ is the collected count the label $C$.
\end{theorem}

To compare the variance between the correlated perturbation and state-of-the-art LDP mechanisms, we use the OUE mechanism as a representative LDP mechanism for item perturbation and GRR for label perturbation~\cite{wang2017locally}.
\begin{theorem}
	Given the dataset $D$ with data amount $N$, item domain $\mathcal{I}$ and class domain $\mathcal{C}$, the label-item pair frequency $f(C, I)$, the class amount $n$, the difference of the variances of the estimated frequency $\hat{f}(C, I)$ in Eq.~(\ref{feoue}) and Eq.~(\ref{unbiased_estimation})  can be derived as 
	\begin{equation}\nonumber
		\begin{aligned}
			V&ar[\hat{f}(C, I)]_{GRR+OUE}-Var[\hat{f}(C, I)]_{CP}>\\
			&\frac{(n-f(C, I))p_1^2q_2^2(1-q_2)^2+(N-n)q_1q_2p_2(1-q_1q_2)^2}{[p_1(1-q_2)(p_2-q_2)]^2}\\
			&+[\frac{q_1q_2(1-p_2)}{p_1(1-q_2)(p_2-q_2)}]^2\frac{np_1(1-p_1)+(N-n)q_1(1-q_1)}{(p_1-q_1)^2}\\
			&+[\frac{q_1}{(p_1-q_1)(p_2-q_2)}]^2[\sum_{C_i\in \mathcal{C} }f(C_i, I)p_2(1-p_2)\\
			&\quad\quad+(N-\sum_{C_i\in \mathcal{C} }f(C_i, I))q_2(1-q_2)].
		\end{aligned}
	\end{equation} 
	\vspace{-1em}
\end{theorem}

Finally, we discuss the utility of the PTS framework with the correlated perturbation and the PTJ framework with an LDP mechanism. Since the PTJ framework consumes the whole privacy budget, integrating OUE into the PTJ framework improves the utility compared with the PTS framework with the correlated perturbation. However, the communication cost in the PTJ framework increases significantly due to the enlarged perturbation domain, which combines all labels and items. Consequently, the PTJ framework is not suitable  for scenarios with a large label domain or item domain due to its high communication cost~\cite{wang2017locally}.
%More details can be found in the experimental parts.
\section{Queries for Multi-class Item Mining}\label{application}
To demonstrate the usability of our framework for multi-class item mining, we implement it for multi-class frequency estimation in Section~\ref{mcfcs} and multi-class top-$k$ item mining in Section~\ref{mctims}, respectively.

\subsection{Multi-class Frequency Estimation}\label{mcfcs}
Many machine learning models, such as the decision tree, rely on frequency information to build models~\cite{li2017feature, quinlan1986induction}. To protect users' private information when building the model, a privacy-preserving approach for multi-class frequency estimation is essential. Building upon the foundational frameworks --- HEC, PTJ, and PTS, we derive the corresponding unbiased frequency estimations within the multi-class context under LDP.

\begin{itemize}
	\item 
	\textbf{Frequency estimation under HEC.} Given a dataset $D$ consisting of $N$ users, with $c$ classes $\mathcal{C}=\{C_1, C_2, \cdots, C_c\}$ and $d$ items $\mathcal{I}=\{I_1, I_2, \cdots, I_d\}$. The users are divided into $c$ groups, each group is responsible for aggregating one class. For each user holding a label-item pair, an LDP mechanism is employed to perturb the item with a privacy budget $\epsilon$.  If a user's label does not match the assigned class, she randomly selects an item from the item domain for perturbation to ensure deniability. The unbiased frequency of a label-item pair $(C, I)$ is then derived as follows: $\hat{f}(C, I) =\frac{c\tilde{f}(C, I)-Nq}{p-q},$
	%\begin{equation}\nonumber
	%		\hat{f}(C, I) =\frac{c\tilde{f}(C, I)-Nq}{p-q},
	%\end{equation}
	where $p$ and $q$ are the corresponding perturbation probabilities of the LDP mechanism.
	
	\item \textbf{Frequency estimation under PTJ.} Given the same dataset illustrated above. Each user perturbs her label-item pair using an LDP mechanism with privacy budget $\epsilon$ over the perturbation domain $\mathcal{P}=\{(C, I)\mid C\in \mathcal{C}, I\in \mathcal{I}\}$. After aggregating all the perturbed pairs, the unbiased frequency for a pair $(C, I)$ can be inferred as: $	\hat{f}(C, I) =\frac{\tilde{f}(C, I)-Nq}{p-q},$
	%\begin{equation}\nonumber
	%	\hat{f}(C, I) =\frac{\tilde{f}(C, I)-Nq}{p-q},
	%\end{equation}
	where $p$ and $q$ are the corresponding perturbation probabilities of the LDP mechanism.

	\item \textbf{Frequency estimation under PTS.} For a label-item pair $(C, I)$ from a user, the label is first perturbed via an LDP mechanism with the privacy budget $\epsilon_1$, and the item is perturbed with the remaining privacy budget $\epsilon_2$. The unbiased estimated frequency is derived in Eq.~(\ref{feoue}).
	
	\item \textbf{Frequency estimation under PTS with the correlated perturbation.} For a user with a label-item pair $(C, I)$, 
	the pair is perturbed via the correlated perturbation with the privacy budget $\epsilon$, which perturbs the item based on the perturbed label.
	The unbiased frequency is derived in Eq.~(\ref{unbiased_estimation}).
\end{itemize}

%\textbf{Communication cost per user.} Given the class domain size $c$ and item domain size $d$, consider the OUE as the LDP mechanism~\cite{wang2017locally}. The communication cost of HEC, PTS, and PTS-CP is $\mathcal{O}(d)$, while for PTJ is $\mathcal{O}(c\cdot d)$ with an enlarged perturbation domain including combinations of both items and labels.
\textbf{Complexity analysis.} Given the class domain size $c$, item domain size $d$, and data amount $N$, we consider the OUE as the LDP mechanism [9]. The communication cost for each user of HEC, PTS, and PTS-CP is $\mathcal{O}(d)$; while for PTJ, it is $\mathcal{O}(cd)$ with an enlarged perturbation domain including combinations of both items and labels. The time complexities of the HEC, PTS, and PTS-CP mechanisms are  $\mathcal{O}(d)$ on the user side, and $\mathcal{O}(Nd)$ on the server side. While the time complexity of PTJ is $\mathcal{O}(cd)$ on the user side and $\mathcal{O}(Ncd)$ on the server side. As for the space complexity, the HEC, PTS, and PTS-CP require $\mathcal{O}(d)$ space on the user side, while the server side requires $\mathcal{O}(cd)$ space. On the other hand, the PTJ requires $\mathcal{O}(cd)$ space on both the user and server sides.

\subsection{Multi-class Top-$k$ Item Mining}\label{mctims}
%Existing top-$k$ mining mechanisms primarily identify the top items, disregarding class information.
%This oversight complicates the perturbation process, as combining labels and items expands the perturbation domain. Furthermore, if a label is perturbed to a different value, it renders the information invalid. 
%To mitigate the effects of label perturbation, correlated perturbation can be utilized. This technique not only manages invalid labels but also leverages invalid items during pruning. Specifically, when infrequent candidates are removed from consideration, users may still possess items that are no longer present in the current candidate set. 
%In this section, we discuss the multi-class top-$k$ item mining.
%Additionally, we refine the mining scheme to improve the effectiveness. 
In this subsection, we first elaborate on two issues in the existing top-$k$ item mining schemes and then propose corresponding solutions. Finally, we present our scheme for multi-class top-$k$ item mining.

\textbf{Existing issues and our solutions.} The top-$k$ item mining is an essential task to identify top frequent items with a large item domain. Many top-$k$ item mining schemes under LDP use a prefix trie to collect item frequencies~\cite{wang2021heavy, wang2018privtrie, bassily2017practical}. 
We take the state-of-the-art method, PEM~\cite{wang2021heavy}, as an example to demonstrate the prefix trie-based schemes under LDP.
Under PEM, items are encoded into bits, converting the top-$k$ item mining problem into a frequent sequence mining task. The trie progressively expands from shallow to deeper levels, as the server collects prefix frequencies and prunes infrequent ones to identify longer frequent prefixes. However, this mining scheme can produce false positive prefixes, and also introduce invalid data during pruning.

\begin{itemize}
	\item \textbf{False positive prefixes}.  While prefix expansion is based on plausible insights, it is not always reliable, as frequent prefixes do not necessarily correspond to frequent sequences~\cite{agrawal1994fast}. The prefix trie may yield false positives, identifying prefixes that appear frequent but are not truly among the top sequences, which potentially causes genuine top items to be overlooked. An example without LDP noise is illustrated in Fig.~\ref{fake}, where the most frequent item `000' is eventually missed as its prefix `0' is less frequent than `1' in the upper layer.
	To mitigate the generation of such false prefixes, a viable approach is to decouple prefix groups to reduce specific combinations, for which shuffling is particularly effective. To reduce the communication cost, only the random seeds and the bucket states for pruning are sent to users rather than the entire set of the shuffled candidates. The shuffling process is illustrated in Fig.~\ref{topkp}.
	\begin{figure}[!htb]
		\vspace{-01em}
		\centering
		\includegraphics[width=0.3\textwidth]{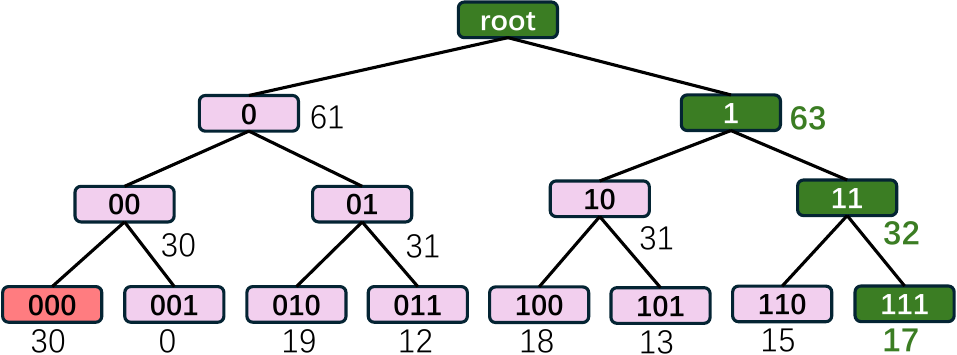}
		\caption{The goal is to mine the top-$1$ item among eight items encoded from `000' to `111', with each node showing the corresponding frequency. With the prefix expansion scheme, the true top item is missed even without LDP noise. Green nodes represent the expansion path of the trie, the red node is the actual top-1 item, and pink nodes are pruned during trie expansion. 
			%			\red{But the mining scheme with shuffling can achieve the top-$1$ item with probability $(C^2_8\cdot C^2_6\cdot C^2_4/4!-C^2_6\cdot C^2_4/3!)/(C^2_8\cdot C^2_6\cdot C^2_4/4!)=0.857$.}
			However, integrating the mining scheme with our shuffling can achieve the top-$1$ item with probability $(C^2_8\cdot C^2_6\cdot C^2_4/4!-C^2_6\cdot C^2_4/3!)/(C^2_8\cdot C^2_6\cdot C^2_4/4!)=0.857$.
		}
		\label{fake}
		\vspace{-1em}
	\end{figure}
	\item\textbf{Invalid data during pruning.} Besides introducing false positive prefixes, existing schemes also neglect the impact of invalid data. As the pruning process progresses, more and more items are excluded from the candidate set, resulting in invalid data. When encountering invalid data, PEM substitutes it with a randomly selected item~\cite{wang2021heavy}. To reduce the impact of invalid data, we apply validity perturbation. 
\end{itemize}

%However, shuffling process incurs a high communication cost, as the server must transmit the shuffled results to each user in a user group responsible for frequency collection. Nonetheless, implementing the shuffling process in real-world scenarios is feasible because it only requires to send random seeds and the remained bucket indices to each user since the item indices is fixed at the beginning. 

\textbf{Multi-class top-$k$ item mining scheme.} For multi-class context, in addition to the shuffling method and validity perturbation, we leverage globally frequent items across classes to enhance utility. We present the details of the PTS-based scheme below with an illustration in Fig.~\ref{topkp}. Note that the PTJ-based scheme can be implemented in a similar way, using the shuffling method and validity perturbation mechanism.
\begin{itemize}
	\item 
	\textbf{Step 1: Candidate generation}. In general, classwise top-$k$ items are also frequent items among the entire dataset. That is because those top frequent items within a class are typically shared among different classes, such as popular goods common to all age groups. Consequently, we can use items from a small group of users sampled from the entire dataset (with parameter $a$ in Algorithm 1 Line 2 controlling the sample proportion) to mine item candidates including these globally frequent items during the initial iterations.  Meanwhile, the perturbed labels can be used to assess the noise level in each class. If the injected noise is too large --- specifically, if it exceeds $b$ times the estimated class size (Algorithm 2 Line 8) --- the amount of valid data may be insufficient for accurate estimation, rendering the correlated perturbation mechanism infeasible. In such a class, only the validity perturbation mechanism is applied. Note that only the PTS framework can benefit from the globally frequent items.  Details are in Algorithm~\ref{gfim}. 
	
	\begin{figure}
		%	\vspace{-1em}
		\begin{algorithm}[H]
			\caption{Candidate generation.}
			\label{gfim}
			\begin{algorithmic}[1]
				\small
				\REQUIRE Dataset $D$, $\epsilon$, $k$, item domain size $d$, class number $|\mathcal{C}|$, a constant $a$
				
				\STATE Obtain the iteration number $IT=\log_2\frac{d}{4k}+1$
				\STATE $\#$ Use $a\cdot \lvert D\rvert$ data to conduct the first $IT_f$ iterations
				\FOR{Iteration $it\le IT_f$}
				\STATE Shuffling and split the item candidates into $4\cdot k\cdot |\mathcal{C}|$ buckets using the given random seeds and bucket states
				\STATE Process each user's item: it is valid if the item is in the pruned candidate set; otherwise, it is invalid.
				\STATE Perturb the item with $\epsilon_2$ via the validity perturbation and perturb the label via an LDP mechanism using $\epsilon_1$
				\STATE Remain top $2\cdot k\cdot |\mathcal{C}|$ buckets as  candidates for next iteration
				\ENDFOR
				\STATE $\#$ Assess the injected noise for each class
				\STATE Estimate user amount $\lvert D'_C\rvert$ for each class from perturbed labels to indicate injected noise levels
				\STATE Return the candidates and noise level for each class
			\end{algorithmic}
		\end{algorithm}
		\vspace{-3em}
	\end{figure}

	\item \textbf{Step 2: Top-$k$ item mining within each class.} After obtaining the candidates and noise level, the remaining users are assigned to each class for classwise top-$k$ item mining based on their perturbed labels. As iterations progress, the proportion of valid data decreases. In the later stages, most of the data becomes invalid due to candidate pruning and label perturbation. To accommodate the decrease of valid data, correlated perturbation is applied only in the final iteration  to identify classwise top frequent items, while the validity perturbation mechanism is adopted in other iterations. The details of this step can be found in Algorithm~\ref{cstim}.
	
	\begin{figure}
		%	\vspace{-1em}
		\begin{algorithm}[H]
			\caption{Top-$k$ item mining within each class.}
			\label{cstim}
			\begin{algorithmic}[1]
				\small
				\REQUIRE Dataset $D_C$, $\epsilon_2$, $k$, the estimated user amount $\lvert D'_C\rvert$, the remaining iteration number $IT_r$, a constant $b$, the candidates from Algorithm~\ref{gfim}
				\STATE $\#$ Process the first $IT_r-1$ iterations
				\FOR{Iteration $it\le IT_r-1$}
				\STATE Shuffling and split the item candidates into $4\cdot k$ buckets with a given random seed
				%			\STATE Process each user's data: it is valid if the held item is in the pruned candidate set; otherwise, it is invalid.
				%			\STATE Perturb the item with $\epsilon/2$ via the validity perturbation and perturb the label via an LDP mechanism using $\epsilon/2$
				\STATE The same steps as Algorithm~\ref{gfim} Line 5 
				\STATE Perturb item with $\epsilon_2$ via the validity perturbation mechanism
				\STATE Remain top $2\cdot k\cdot$ buckets as  pruned items for next iteration
				\ENDFOR
				\STATE $\#$ Process the last iteration
				\STATE If the collected user amount $\lvert D_C\rvert>b\cdot \lvert D'_C\rvert$, only apply validity perturbation mechanism in the last iteration
				\STATE Choose the mechanism from the validity perturbation mechanism and correlated perturbation mechanism according to the noise level
				\STATE Return the top-$k$ items according to the aggregation
			\end{algorithmic}
		\end{algorithm}
		\vspace{-2em}
	\end{figure}
\end{itemize}

%\textbf{Communication cost per user.} Given the class domain size $c$ and item domain size $d$, let's consider the OUE as an example of the LDP mechanism for item perturbation~\cite{wang2017locally}. The communication costs of HEC, PTS under the PEM scheme, and the PTS with optimized methods are $\mathcal{O}(k \log d)$. For PTJ under the PEM scheme and the PTJ with optimized methods, the communication costs are $\mathcal{O}(ck \log c \cdot d)$. The communication cost for PTJ is the highest due to its enlarged perturbation domain, which includes combinations of both items and labels.
\textbf{Complexity analysis.} Given the class domain size $c$, item domain size $d$, and data amount $N$, we use the GRR mechanism for label perturbation and the OUE mechanism for item perturbation [9], the communication cost is measured for each user. The mining scheme for the fundamental frameworks is PEM, and the extending length in each iteration is $m$~\cite{wang2017locally}. The costs are summarized in Table ~\ref{tabc}.
%	The communication costs of HEC, PTS under the PEM scheme are $\mathcal{O}(2^mk \log d)$, while for the PTS with optimized methods are $\mathcal{O}(k \log d)$. For PTJ under the PEM scheme, the communication cost is $\mathcal{O}(2^mck \log cd)$, whereas for PTJ with optimized methods is $\mathcal{O}(ck \log cd)$. The communication cost for PTJ is the highest due to its enlarged perturbation domain, which includes combinations of both items and labels. The time complexity of HEC is $\mathcal{O}(2^mk)$ for the user side, while the server side is $\mathcal{O}(2^mck(m+\log k)\frac{\log d}{m})$. The time complexity of PTJ is $\mathcal{O}(2^mck)$, while the server side is $\mathcal{O}(2^mck(m+\log ck)\frac{\log cd}{m})$. The time complexity The time complexity of PTS with optimized methods is $\mathcal{O}(ck)$, while the server side is $\mathcal{O}(2^mck(m+\log ck)\frac{\log cd}{m})$.

\begin{table}[h]
	\vspace{-1em}
	\centering
	\scriptsize
	\caption{Complexity Analysis}
	\label{tabc}
	
	\begin{tabular}{p{0.5cm}| p{1.7cm} p{3.8cm} p{1.5cm}}  % l=左对齐, c=居中, r=右对齐
		\hline
		\textbf{} & \textbf{Communication} & \textbf{Time}& \textbf{Space}  \\ 
		\hline
		\multirow{2}{*}{\parbox {0.5cm}{\textbf{HEC}\\\textbf{PTS}}} & \multirow{2}{*}{$\mathcal{O}(2^mk \log d)$} & $\mathcal{O}(2^mk)$  & $\mathcal{O}(2^mk\log d)$ \\ 
		&                      & $\mathcal{O}(2^mk[c(m+\log k)\frac{\log d}{m}+N])$ & $\mathcal{O}(2^mck\log d)$  \\ \hline
		\multirow{2}{*}{\textbf{PTJ}} & \multirow{2}{*}{$\mathcal{O}(2^mck \log cd)$} & $\mathcal{O}(2^mck)$  & $\mathcal{O}(2^mck\log cd)$ \\ 
		&                      & $\mathcal{O}(2^mck[(m+\log ck)\frac{\log cd}{m}+N])$ & $\mathcal{O}(2^mck\log cd)$  \\ \hline
		\multirow{2}{*}{\textbf{PTJ$^\dagger$}} & \multirow{2}{*}{$\mathcal{O}(ck)$} & $\mathcal{O}(ck)$  & $\mathcal{O}(cd)$ \\ 
		&                      & $\mathcal{O}(ck(\log ck \log \frac{d}{k}+N))$ & $\mathcal{O}(cd)$  \\ \hline
		\multirow{2}{*}{\textbf{PTS$^\dagger$}} & \multirow{2}{*}{$\mathcal{O}(ck)$} & $\mathcal{O}(ck)$  & $\mathcal{O}(d)$ \\ 
		&                      & $\mathcal{O}(ck(\log ck \log \frac{d}{k}+N))$ & $\mathcal{O}(cd)$  \\ \hline
	\end{tabular}
	
	\vspace{0.15em}
	\scriptsize{$^\dagger$  represent the optimized methods, with the first line of each row showing the user-side results and the second line showing the server-side results.}
	\vspace{-2em}
\end{table}

\begin{figure*}[btp]
	\centering
	\includegraphics[width=0.8\textwidth]{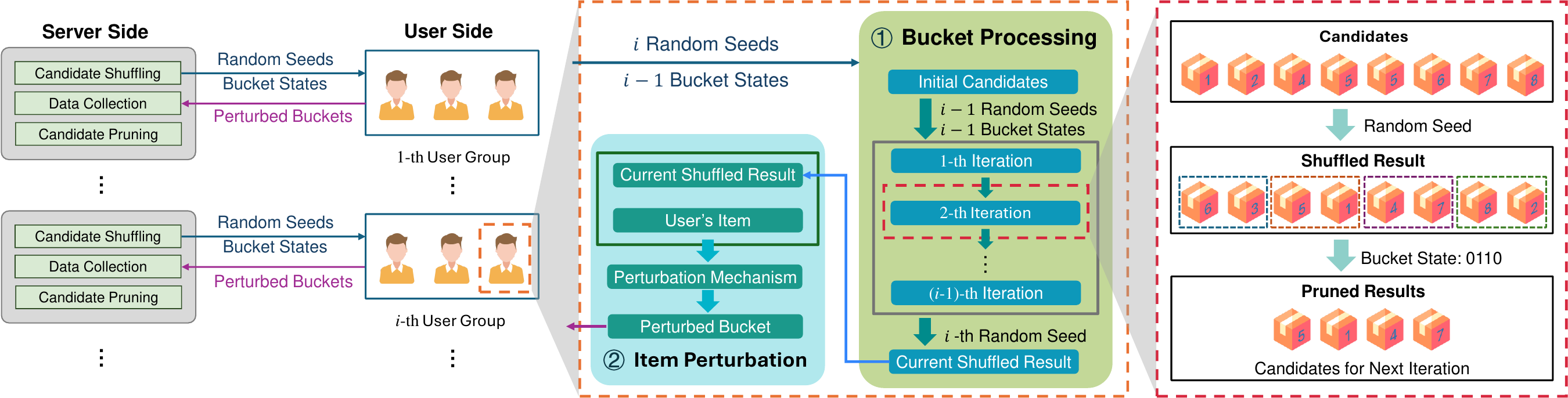}
	\caption{Each user receives random seeds and bucket states to generate a current shuffled result before perturbing her label-item pair. The choice of perturbation mechanism depends on the aggregation goal described in Algorithms~\ref{gfim} and~\ref{cstim}.}
	\label{topkp}
	\vspace{-1.5em}
\end{figure*}

\section{Experimental Evaluation}\label{experiments}
In this section, we empirically evaluate the performance of the three frameworks, HEC, PTJ, and PTS, and the optimized methods for item mining tasks. All methods are implemented in Python 3, and each experiment is averaged from 20 trials\textsuperscript{$\ddagger$}.\footnote{$\ddagger$ Our code is at  github.com/Abigail-MAO/Multi-class-Item-Mining}
%We compare the performance of the three frameworks: HEC, PTJ, and PTS. For the perturbation in PTJ and HEC, we select the mechanism based on Wang et al.'s conclusion, choosing between Generalized Random Response (GRR) and Optimal Unary Encoding (OUE).
%To ensure a fair comparison, we implement the PTS framework with the OUE mechanism for item perturbation~\cite{wang2017locally}.

%conduct experiments on a server cluster with twenty dual-Intel(R) Xeon(R) CPU E5-2640 v3 @ 2.60GHz, 64G RAM, Ubuntu 20.04 LTS operating system.

\subsection{Datasets}
We conduct the experiments over six datasets including both real-world datasets and synthetic datasets.

%\noindent\textbf{Multi-class Frequency Estimation.} We use two machine learning datasets with multiple features and labels. To align with our problem setting, where each user holds a label-item pair, we calculate item frequency by dividing users into groups, with each group focused on a single feature.
%\begin{itemize}
%	\item \textbf{Comprehensive Diabetes Clinical Dataset.}\footnote{https://www.kaggle.com/datasets/iammustafatz/diabetes-prediction-dataset} This dataset is collected from 100,000 individuals for diabetes-related research. The dataset contains eight features including categorical values like gender and continuous values like blood glucose level. For the continuous values, we round to one decimal place and the largest domain contains approximately 600 items.
%	\item \textbf{Heart Disease Health Indicator Dataset.}\footnote{https://www.kaggle.com/datasets/alexteboul/heart-disease-health-indicators-dataset} This dataset contains 253,680 survey responses from cleaned BRFSS 2015 to be used primarily for the binary classification of heart disease. The dataset contains 21 features and each feature only contains categorical values. And the largest item domain is 84 in this dataset.
%\end{itemize}

\begin{itemize}
	\item \textbf{Comprehensive Diabetes Clinical Dataset.}\textsuperscript{$^1$}\footnote{$^1$kaggle.com/datasets/iammustafatz/diabetes-prediction-dataset} It is collected from 100,000 individuals for diabetes research, including eight features. And continuous values are rounded to one decimal place, with the largest feature domain containing about 600 items.
	
	\item \textbf{Heart Disease Health Indicator Dataset.}\textsuperscript{$^2$}\footnote{$^2$kaggle.com/datasets/alexteboul/heart-disease-health-indicators-dataset} This dataset includes 253,680 cleaned survey responses from BRFSS 2015, primarily for binary heart disease classification. It contains 21 categorical features, with the largest item domain being 84.
	
	\item \textbf{MyAnimeList Dataset.}\textsuperscript{$^3$}\footnote{$^3$kaggle.com/datasets/azathoth42/myanimelist}
	This dataset contains information on anime viewing habits, comprising around 116,000 users, 14,000 anime titles, and 35 million records. We treat gender and watched anime as label-item pairs, mining the top anime titles across different gender groups. We sample 20\% data for experiments.
	
	\item \textbf{JD Contest Dataset.}\textsuperscript{$^4$}\footnote{$^4$kaggle.com/datasets/pwang001/jjingdong-contest-dataset} This dataset from JD.COM contains sale records about 28,000 items. We utilize the five age groups (below 25, 26-35, 36-45, 46-55, and above 56) and treat the age group as the label, resulting in 45 million valid item-label pairs. A 20\% sample is used for experiments.
	
	\item \textbf{Synthetic Datasets.} We generate four datasets: SYN1 and SYN2 are used for variance analysis, while SYN3 and SYN4 are employed to study the effect of varying class numbers, ranging from 10 to 50. To control $f(C, I)$, $n$, and $N$, SYN1 and SYN2 are generated with four classes, four items, and label-item pair amounts of 1,000, 10,000, 100,000, and 1,000,000. SYN1 fixes the class amount $n$  to investigate correlation strength varying, while SYN2 fixes one label-item pair frequency $f(C, I)$ across the classes to examine the impact of the class distribution.  SYN3 and SYN4 contain 20,000 items and five million instances, the data size of each class satisfies the normal distribution. Within each class, the items are drawn from the exponential distribution with the scale from 0.01 to 0.1. SYN3 is simulated based on real-world datasets and includes globally frequent items,  with an average of eight overlapping items among the top 20 items between any two classes. In contrast, SYN4 is generated using the same method but excludes globally frequent items.
	
	%	SYN1 has 50 classes and 20,000 items, with a fixed overall distribution as class numbers change. The class sizes follow a normal distribution, ranging from 500,000 to 1,000,000 instances, and items are drawn from an exponential distribution (scale 0.1 to 0.01). To simulate real-world item distributions, we adjust the overlap of top items between classes, with an average of eight overlapping items in the top-20 across classes. For each class number configuration, each class data is randomly sampled from the 50 classes. SYN2 is similar, but with a fixed total of 5,000,000 instances.
	
\end{itemize}

We use the Diabetes and Heart Disease datasets for the frequency estimation task. To align with our problem setting where each user holds a label-item pair, we calculate label-item frequency within each feature by dividing users into groups, with each group focusing on a single feature to mine corresponding label-item pairs. Moreover, we assess the performance of top-$k$ item mining using the Anime and JD datasets, while the two synthetic datasets are used to analyze the impact of varying class numbers.

\subsection{Metrics}
%We use different metrics to assess the performance of frequency estimation and top-$k$ item mining.

For frequency estimation, we apply root mean square error (RMSE) to measure the difference between the estimated frequency $\hat{f}(C, I)$ of label-item pairs and the ground truth $f(C, I)$,
\begin{equation}\nonumber
	RMSE=\sqrt{\frac{1}{\lvert\mathcal{C}\rvert\cdot\lvert\mathcal{I}\rvert}\sum_{C\in\mathcal{C}}\sum_{I\in\mathcal{I}}(\hat{f}(C, I)-f(C, I))^2},
\end{equation}
where $\mathcal{C}$ is the class domain, $\mathcal{I}$ is the item domain, and $\lvert\cdot\rvert$ denotes the set size.

To compare the mined top-$k$ items $\mathcal{I}^m$ with the ground truth $\mathcal{I}^g=\{I^g_1, I_2^g, \cdots, I_k^g\}$ within each class, we use the same metrics as in PEM~\cite{wang2021heavy}: F1 Score~\cite{manning2008introduction} and Normalized Cumulative Rank (NCR)~\cite{jarvelin2002cumulated}. Notably, since precision equals recall in this context, the F1 Score evaluates the ratio of mined true positive items. The NCR measures the quality of the mined items and is defined as $NCR=\frac{2\sum_{I_i\in \mathcal{I}^m}q(I_i)}{k(k+1)},$
%\begin{equation}\nonumber
%	NCR=\frac{2\sum_{I_i\in \mathcal{I}^m}q(I_i)}{k(k+1)},
%\end{equation}
where $q(I_1^g)=k, q(I_2^g)=k-1, \cdots, q(I_k^g)=1$. To obtain an overall assessment, we average the F1 Score and NCR across the classes.

\subsection{Variance Analysis}
\begin{figure}[htbp]
	\centering  %图片全局居中
	
	\includegraphics[width=0.11\textwidth]{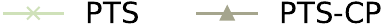}
	
	\subfigure[Varying correlation strength.]{
		\includegraphics[width=0.2\textwidth]{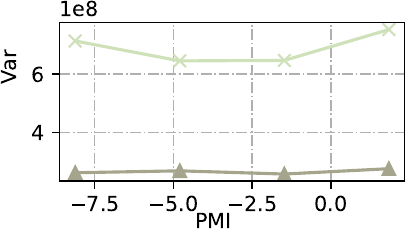}\label{var_a}}
	\hspace{0.5em}
	\subfigure[Varying class amount $n$.]{
		\includegraphics[width=0.2\textwidth]{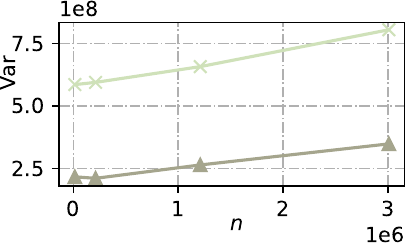}\label{var_b}}
	\vspace{-0.5em}
	\caption{Empirical variance analysis.}
	\label{var}
	\vspace{-1em}
\end{figure}

To investigate the impact of label-item correlations on variance, we conduct experiments on the SYN1 dataset. We fix item amount and class amount ($f(I)=n=1.111\times 10^6$), and vary label-item frequencies ($f(C, I)\in\{10^3, 10^4, 10^5, 10^6\}$) in each class. The variance is computed as $Var[\hat{f}(C, I)]=\frac{1}{t}[\hat{f}(C, I)-f(C, I)]^2$~\cite{wang2017locally} with $\epsilon=1$ and experimental time $t=1000$. And correlation strength measure PMI~\cite{church1990word} is calculated according to $f(C, I)$ within one class. PTS refers to the framework PTS with GRR and OUE, while PTS-CP denotes the PTS framework with our improved mechanism, the correlated perturbation (CP).  As shown in Fig.~\ref{var_a}, despite the increase in PMI, the observed variance exhibits negligible variation, empirically confirming that  changes in correlation strength are concealed in variance due to the dominance of class amount $n$ and data amount $N$. To examine class distribution effects, we utilize the SYN2 dataset with fixed label-item frequency $f(C, I)=10^4$ for an item across the classes, and varying class amount $n\in\{1.3\times 10^4, 2.11\times 10^5, 1.21\times 10^6, 3.01\times 10^6\}$. As shown in Fig.~\ref{var_b}, the results demonstrate a strong positive correlation between class amount $n$ and variance magnitude. These quantitative relationships directly validate the theoretical analysis derived in Section V-C.
\vspace{-1em}

\subsection{Results of Frequency Estimation}
\begin{figure}[htbp]
	\centering  %图片全局居中
	\vspace{-1em}
	\includegraphics[width=0.22\textwidth]{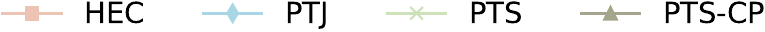}
	
	\subfigure[Diabetes Dataset.]{
		\includegraphics[width=0.18\textwidth]{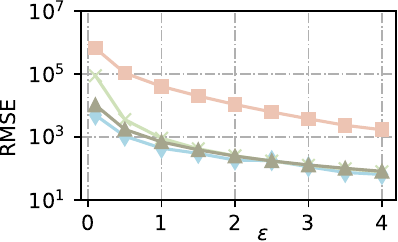}}
	\hspace{1em}
	\subfigure[Heart Disease Dataset.]{
		\includegraphics[width=0.18\textwidth]{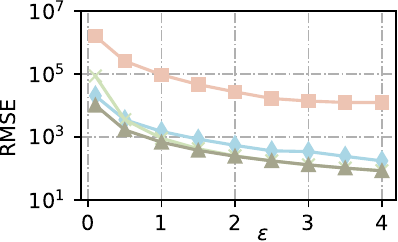}}
	\vspace{-0.5em}
	\caption{RMSE from two real-world datasets with varying privacy budget $\epsilon$.}
	\label{fprw}
	\vspace{-0.5em}
\end{figure}

For frequency estimation, the multi-class item mining frameworks incorporate the state-of-the-art mechanism, the adaptive mechanism, which adaptively runs  OUE and GRR according to the item domain size $d$~\cite{wang2017locally}. Specifically, if $d>3e^\epsilon+2$, OUE is employed to decrease the variance. The HEC and PTJ frameworks use the adaptive mechanism. In the PTS framework, where the label domain is small and the item domain is large, GRR is applied for label perturbation, while OUE is used for item perturbation. Since PTJ does not produce any invalid data in this task, the correlated perturbation can only be integrated with the PTS framework (designated as PTS-CP), where the privacy budgets are set as $\epsilon_1=\epsilon_2=\epsilon/2$. The details have been derived in Section~\ref{mcfcs}. Comparative results are presented in Fig.~\ref{fprw}.

As shown in Fig.~\ref{fprw}, PTJ and PTS significantly outperform HEC. Correlated perturbation (PTS-CP) notably enhances utility compared with the PTS framework, particularly when the privacy budget $\epsilon$ is small. As the privacy budget increases, label perturbation results in fewer instances of invalid data, further improving the utility of PTS. PTJ, in particular, performs the best on the Diabetes dataset, benefiting from the utilization of the whole privacy budget and the OUE mechanism~\cite{wang2017locally}. However, despite this advantage, PTJ incurs the highest communication load due to its expanded perturbation domain that combines both label and item domains.

\subsection{Results of Top-$k$ Item Mining}
In this section, we present the results of top-$k$ item mining. For comparison, we employ the state-of-the-art top-$k$ item mining scheme, PEM~\cite{wang2021heavy}, on multi-class item mining frameworks, compared with our optimized methods, including validity perturbation (VP), correlated perturbation with the globally frequent items (CP), and the shuffling method tailored for top-$k$ mining scheme (Shuffling). As for the parameters, we empirically set $\epsilon_1=\epsilon_2=\epsilon/2$, and $IT_f=int(IT/2)$ in Algorithms~\ref{gfim} and~\ref{cstim}. We choose $a=0.2$, meaning that one-fifth of the data will be used for generating global candidates, while the remaining data is used for top-$k$ item mining. The parameter $b$ is set to 2, meaning that correlated perturbation is only applied when the collected class amount is less than twice the estimated class amount. We examine the impact of varying the privacy budgets, $k$ values, and class numbers, along with an ablation study of the optimized methods.

\begin{figure}[htbp]
	\centering  %图片全局居中
	\includegraphics[width=0.4\textwidth]{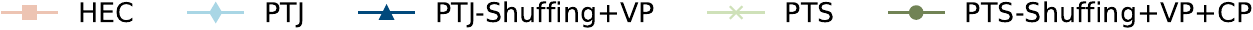}
	
	\subfigure[Anime dataset]{
		\includegraphics[width=0.18\textwidth]{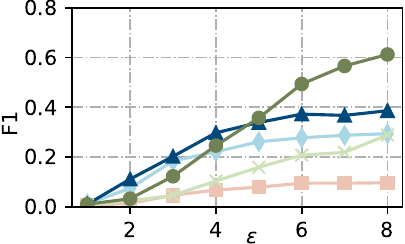}}
	\hspace{1em}
	\subfigure[Anime dataset]{
		\includegraphics[width=0.18\textwidth]{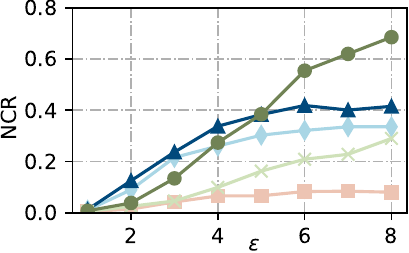}}
	
	\vspace{-1em}
	\subfigure[JD dataset]{
		\includegraphics[width=0.18\textwidth]{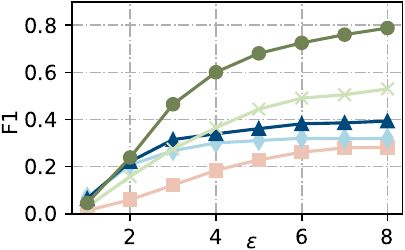}}
	\hspace{1em}
	\subfigure[JD dataset]{
		\includegraphics[width=0.18\textwidth]{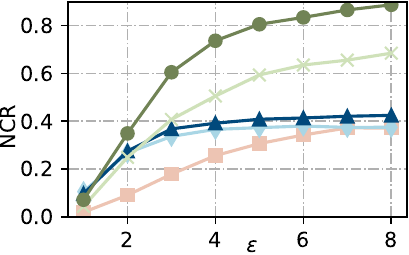}}
	\vspace{-0.5em}
	\caption{Results from real-world datasets with $k=20$ and varying  $\epsilon$.}
	\label{topkev}
	\vspace{-1em}
\end{figure}

The results of epsilon varying are shown in Fig.~\ref{topkev}, with $k$ fixed as 20.   All methods improve as the privacy budget increases, with PTS-based methods demonstrating a more significant enhancement. Moreover, the optimized methods consistently outperform the initial frameworks. For the Anime dataset, the optimized methods on the PTS framework achieve average improvements of 116.6\% in F1 Score and 134.3\% in NCR, while the optimized methods on the PTJ framework yield average improvements of  25.6\% in F1 Score and 25.0\% in NCR Score. In the JD dataset, the optimized methods on the PTS framework achieve average improvements of 55.6\% in F1 Score and 40.8\% in NCR Score, whereas the optimized methods on the PTJ framework enhance the F1 Score by 13.6\% and the NCR by 6.9\% on average.
\vspace{-1em}

\begin{figure}[htb]
	\centering  %图片全局居中
	\includegraphics[width=0.4\textwidth]{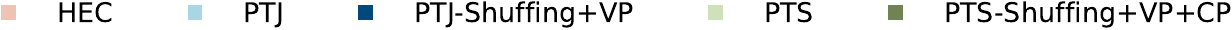}
	\subfigure{
		\includegraphics[width=0.25\textwidth]{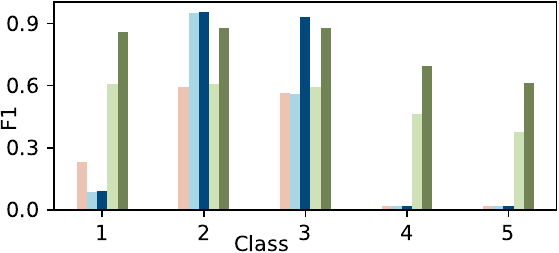}}
	%	\hspace{1em}
	%	\subfigure{
		%		\includegraphics[width=0.2\textwidth]{figures/7-exp/category/ncg_score-crop.pdf}}
	\vspace{-1em}
	\caption{Results of each class on JD dataset with $\epsilon=8$ and $k=20$.}
	\label{eachclass}
	\vspace{-1.2em}
\end{figure}

We also investigate the performance of each class. For the sake of brevity, we only present the F1 Score on the JD dataset with $\epsilon=8$ and $k=20$ in Fig.~\ref{eachclass}, as the NCR Score follows a similar trend. The results match the data size in each class, with instance counts of 850k, 4m, 3m, 314k, and 170k, respectively. Classes 2 and 3 contain significantly more data than the others, and the data volumes in classes 4 and 5 are insufficient to yield reliable results. The PTS framework can benefit from the globally frequent items, even those items with falsely perturbed labels from other classes. Conversely, PTJ can not utilize the global information, and thus fails to produce results in classes 4 and 5.
\vspace{-1em}

\begin{figure}[htb]
	\centering  %图片全局居中
	\includegraphics[width=0.4\textwidth]{figures/epsilon_varying_legend-crop.pdf}
	\subfigure{
		\includegraphics[width=0.2\textwidth]{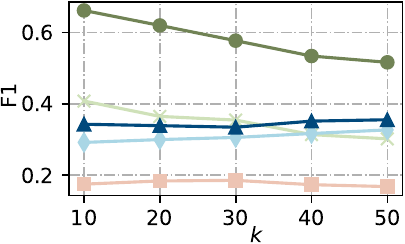}}
	\hspace{1em}
	\subfigure{
		\includegraphics[width=0.2\textwidth]{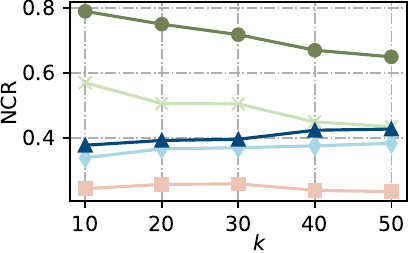}}
	\vspace{-0.8em}
	\caption{F1 Score and NCR on JD dataset with $\epsilon=4$ and varying $k$.}
	\label{varyingk}
	%\vspace{-0.5em}
\end{figure}

In addition to examining the impact of varying privacy budgets, we also explore the influence of the top-$k$ setting, with $k$ ranging from 10 to 50, and the privacy budget is set as $\epsilon=4$. For brevity, we present the results of the JD dataset in Fig.~\ref{varyingk}. As $k$ increases, the utility of PTS decreases, as less frequent items become harder to detect. In contrast, the utility of PTJ improves since a larger $k$ results in a much larger candidate set, allowing more candidate label-item pairs to be investigated.
\vspace{-1em}

\begin{figure}[!h]
	\vspace{-0.6em}
	\centering  %图片全局居中
	\includegraphics[width=0.4\textwidth]{figures/epsilon_varying_legend-crop.pdf}
	
	\subfigure[SYN3 dataset (Global)]{
		\includegraphics[width=0.2\textwidth]{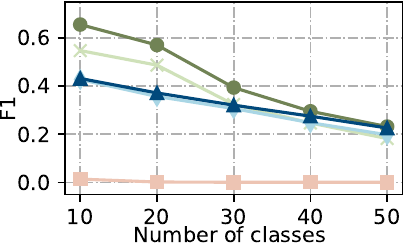}}
	\hspace{1em}
	\subfigure[SYN3 dataset (Global)]{
		\includegraphics[width=0.2\textwidth]{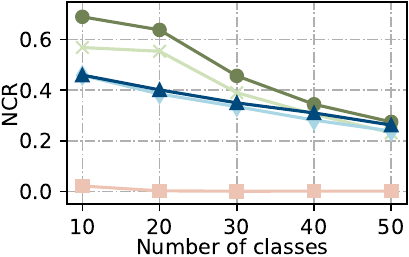}}
	
	\vspace{-0.2em}
	\subfigure[SYN4 dataset]{
		\includegraphics[width=0.2\textwidth]{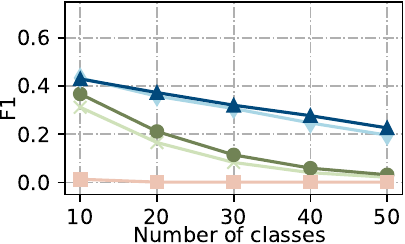}}
	\hspace{1em}
	\subfigure[SYN4 dataset]{
		\includegraphics[width=0.2\textwidth]{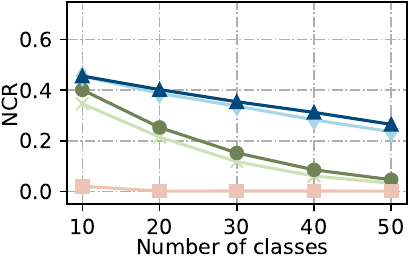}}
	\vspace{-0.8em}
	\caption{Synthetic datasets with $\epsilon=4$, $k=20$, and varying class numbers.}
	\label{varyingclass}
	\vspace{-0.8em}
\end{figure}

\vspace{-0.5em}
Furthermore, we examine the impact of varying class numbers using two synthetic datasets, SYN3 and SYN4. SYN3 contains globally frequent items, whereas SYN4 is generated using the same method but excludes these items. The task is to identify the top 20 items with a privacy budget of $\epsilon=4$. The results are shown in Fig.~\ref{varyingclass}.  Notably, the utility of all methods declines as the class number increases, and all optimized methods perform better than the original frameworks. Without access to globally frequent items, the utility of PTS degrades significantly in SYN4. In contrast, the results of PTJ on both synthetic datasets are similar, indicating that the PTJ framework does not benefit from the inclusion of globally frequent items. Moreover, the PTJ framework suffers from higher communication costs due to the expanded domain size.

To further analyze the impact of the optimized methods, we conduct an ablation study, fixing $k=20$ and $\epsilon=5$. For brevity, we present only the results from the Anime dataset. As shown in Table ~\ref{abalationPTJ}, all optimized methods demonstrate effectiveness, with improvements being particularly pronounced in the PTS framework. This enhancement in PTS stems from its ability to leverage global information across classes with correlated perturbation. Although HEC can also utilize global information, it lacks the ability to differentiate class-specific details at the final stage. We also observe that the shuffling method and validity perturbation enhance both the PTS and PTJ frameworks effectively.
\begin{table}[htbp]
	\vspace{-0.8em}
	\renewcommand{\arraystretch}{1.5}
	\centering
	\vspace{-1em}
	\caption{ablation study on PTJ and PTS}
	\vspace{-0.5em}
	\begin{tabular}{|>{\centering\arraybackslash}m{0.5cm}|>{\centering\arraybackslash}m{1cm}|>{\centering\arraybackslash}m{0.5cm}|>{\centering\arraybackslash}m{1cm}|>{\centering\arraybackslash}m{2.2cm}|}
		\hline
		\textbf{PTJ+} & \textbf{(Baseline)} & \textbf{VP} & \textbf{Shuffling} & \textbf{All optimizations} \\ \hline
		\textbf{F1} & 0.261  & 0.280 & 0.316 & 0.340 \\ \hline
		\textbf{NCR} & 0.303  & 0.326 & 0.360 & 0.387 \\ \hline
	\end{tabular}
	
	\vspace{1em}
	
	\begin{tabular}{|>{\centering\arraybackslash}m{0.5cm}|>{\centering\arraybackslash}m{1cm}|>{\centering\arraybackslash}m{1cm}|>{\centering\arraybackslash}m{0.5cm}|>{\centering\arraybackslash}m{1cm}|>{\centering\arraybackslash}m{2.2cm}|}
		\hline
		\textbf{PTS+} & \textbf{(Baseline)} & \textbf{Global} & \textbf{VP} & \textbf{Shuffling} & \textbf{All optimizations} \\ \hline
		\textbf{F1} & 0.159 & 0.165 & 0.214 & 0.241 & 0.358 \\ \hline
		\textbf{NCR} & 0.163 & 0.180 & 0.229 & 0.270 & 0.385 \\ \hline
	\end{tabular}
	\label{abalationPTJ}
	\vspace{-2em}
\end{table}

\subsection{Impacts of Parameters}
\begin{figure}[!h]
	\centering  %图片全局居中
	\subfigure[5 Classes]{
		\includegraphics[width=0.15\textwidth]{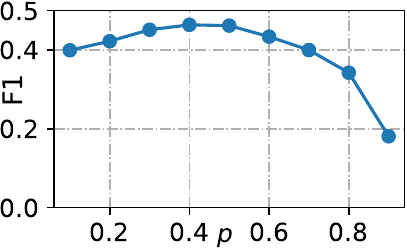}}
	\subfigure[10 Classes]{
		\includegraphics[width=0.15\textwidth]{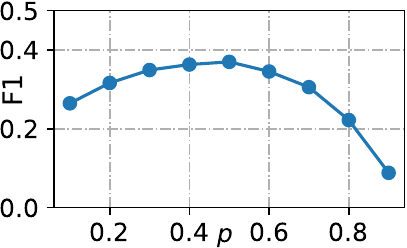}}
	\subfigure[20 Classes]{
		\includegraphics[width=0.15\textwidth]{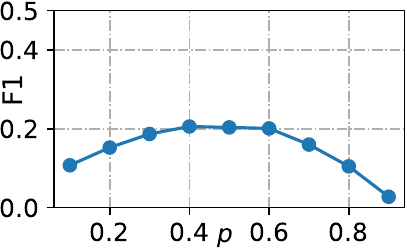}}
	\vspace{-0.8em}
	\caption{Varying the proportion of privacy budget allocation.}
	\label{pba}
	\vspace{-1em}
\end{figure}

In this section, we investigate the impact of parameters, including the settings of privacy budget allocation, and the parameters $a$ and $b$ in Algorithms 1 and 2. In the interest of space, we only present the results of the F1 score, as the NCR follows a similar trend. As aforementioned, the privacy budget is divided into two parts: $\epsilon_1$ for label perturbation and $\epsilon_2$ for item perturbation. As both item and label perturbation affect the aggregation of class-specific items, the perturbations for both are crucial. To investigate the impact of privacy budget allocation, we conduct experiments on the synthetic dataset SNY4 with 5, 10, and 20 classes, respectively. The proportion of $\epsilon_1$, denoted by $p$, varies from 0.1 to 0.9. The results are shown in Fig. \ref{pba}. We observe that the F1 Score increases and then decreases with $p$, which is consistent with our analysis as above. The best $p$ lies between 0.4 and 0.6, and does not influence the results significantly. Therefore, we empirically set the proportion to 0.5 in our paper, i.e., $\epsilon_1=\epsilon_2=\epsilon/2$.

\begin{figure}[!h]
	\vspace{-0.8em}
	\centering  %图片全局居中
	\subfigure[Anime Dataset]{
		\includegraphics[width=0.17\textwidth]{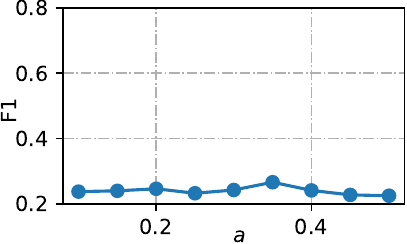}\label{aa}}\hspace{1em}
	\subfigure[JD Dataset]{
		\includegraphics[width=0.17\textwidth]{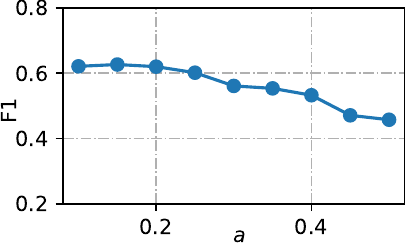}\label{aj}}
	
	\vspace{-0.5em}
	\subfigure[Anime Dataset]{
		\includegraphics[width=0.17\textwidth]{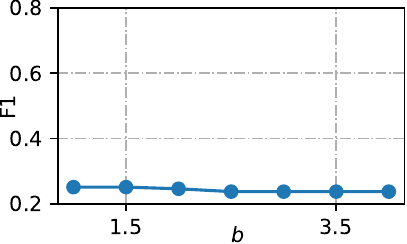}\label{ba}}\hspace{1em}
	\subfigure[JD Dataset]{
		\includegraphics[width=0.17\textwidth]{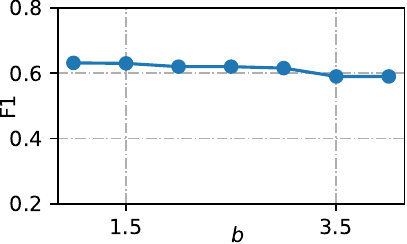}\label{bj}}
	\vspace{-0.8em}
	\caption{Varying the parameters $a$ and $b$.}
	% \label{pba}
	\vspace{-1em}
\end{figure}

Parameter $a$ (Algorithm 1 Line 2) controls the sampled group size used to identify item candidates, including the globally frequent items during the initial iterations. Given a large value of $a$, there may be insufficient data for top-$k$ item mining, while for a small value of $a$, it may fail to generate reasonable candidates. To figure out the impact of parameter $a$, we conduct experiments using varying values of $a$ on the real-world datasets JD and Anime. The results are shown in Figs. \ref{aa} and \ref{aj}.  Clearly, the impact of parameter $a$ depends on the distribution of the dataset. For simplicity, we choose $a=0.2$  in our experiments, meaning that one-fifth of the data is sampled to generate the global candidates, while the remaining data is used for top-$k$ item mining. We investigate the parameter $b$ (Algorithm 2 Line 8) in a similar manner. The data amount within different classes varies a lot. After applying DP perturbations, classes with less data are injected with a significant amount of noise (i.e., items) from other classes. To estimate the noise level for each class, we use the labels perturbed during the global candidates aggregation. If the injected noise is too large (or specifically, the collected data amount exceeds $b$ times the estimated class data amount), the noise level becomes excessive. In this case, the amount of valid data may be insufficient for accurate estimation, rendering the correlated perturbation mechanism infeasible. To study the impact of parameter $b$, we conduct experiments with varying $b$ on the real-world datasets JD and Anime. The results are shown in Figs.\ref{ba} and \ref{bj}. Although the results are dependent on the dataset, they do not fluctuate significantly. In our experiments, we choose $b=2$ as the default value.

\vspace{-0.1em}
\section{Related Works}\label{relatedworks}
\vspace{-0.1em}
In this section, we review existing works on local differential privacy, with a focus on frequency estimation and top-$k$ item mining.

\textbf{Local differential privacy and its applications.} Differential privacy was introduced as a privacy-preserving technique with a theoretical guarantee for data collection~\cite{dwork2006differential}. However, differential privacy requires a trusted third party to aggregate data from users which is not feasible in many real-world scenarios. To this end, local differential privacy (LDP) is proposed~\cite{ye2020local, kasiviswanathan2011can}. In the context of LDP, users will perturb their data locally before uploading it to the untrusted server. Since the advent of LDP, it has been widely applied to various statistical collection tasks, including mean and frequency estimations~\cite{wang2017locally, li2020estimating, erlingsson2014rappor, wang2019collecting}.

\textbf{Local differential privacy item mining.} To meet the growing demand for item mining, researchers have proposed various mechanisms for frequency estimation and top-$k$ item mining. RAPPOR was developed to aggregate frequencies of categorical input items~\cite{erlingsson2014rappor}.  Subsequently, Optimal Unary Encoding (OUE) and Optimal Local Hashing (OLH) were introduced to achieve unbiased frequency estimation with optimal variance~\cite{wang2017locally}. These two mechanisms remain state-of-the-art for frequency estimation, with OUE being more commonly adopted due to its ease of implementation, finding application in various downstream tasks~\cite{du2023ldptrace, huang2024ldpguard, hu2024real}. In terms of top-$k$ item mining, tree-based data structures are often preferred due to their efficiency in performing multi-iteration pruning and estimation~\cite{bassily2017practical, wang2021heavy, wang2018privtrie}. The PEM algorithm~\cite{wang2021heavy}, recognized as a state-of-the-art method, is designed to mine top-$k$ items from large categorical datasets. Building on PEM, Zhu et al.\cite{zhu2023heavy} proposed additional mechanisms to enhance top-$k$ mining in set-valued datasets. Recent advancements include Du et al.'s~\cite{du2024top} adaptive sampling technique for set-valued data to enhance utility, and Li et al.'s~\cite{li2024local} application of the HeavyGuardian data structure for mining top-$k$ items in bounded-memory data streams.

%Note that the aforementioned item mining mechanisms focus exclusively on global statistics collected from the entire dataset. However, incorporating class information can enhance utility in many real-world applications, such as recommendation systems. Thus far, only a variant of differential privacy, known as label differential privacy (label-DP), has been proposed to introduce label protection~\cite{ghazi2021deep}.
% Label-DP assumes that feature information is non-sensitive, requiring only labels to be protected. This assumption is applicable only in specific scenarios, which limits the broader applicability of label-DP~\cite{malek2021antipodes}.
The aforementioned item mining mechanisms rely exclusively on global statistics from the entire dataset, without incorporating class information. While integrating class information can benefit real-world applications such as recommendation systems, it requires a higher privacy budget to ensure privacy guarantees. To date, only label differential privacy (label-DP) has been proposed to protect labels~\cite{ghazi2021deep}, but it does not extend protection to other sensitive information. Label-DP assumes that feature information is non-sensitive, focusing solely on label protection --- an assumption that limits its applicability to specific scenarios~\cite{malek2021antipodes}. In multi-class item mining settings, where both items and labels are sensitive, label-DP fails to meet the necessary privacy requirements.

To the best of our knowledge, we are the first to investigate multi-class item mining under LDP, proposing foundational frameworks and corresponding optimized mechanisms to enhance utility.

\vspace{-0.5em}
\section{Conclusion and Future Work}\label{conclusion}
In this paper, we propose two frameworks, PTJ and PTS, for multi-class item mining, accompanied by two optimized mechanisms as the perturbation module: validity perturbation mechanism and correlated perturbation mechanism. These optimized methods are applied to two types of item mining tasks: frequency estimation and top-$k$ item mining in the multi-class setting. Our core idea is to handle invalid data while preserving the relationship between labels and items. Additionally, we derive unbiased frequency estimation and optimize the top-$k$ item mining scheme. Theoretical analysis and experimental results validate the effectiveness and superiority of the proposed mechanisms.

As for future work, we aim to study multi-class item mining on more data types, such as numerical items. Additionally, we plan to extend this study to real-world mining applications, including gradient descent optimization and k-means clustering.

%\textbf{Limitation and future work.} Our framework relies on the Optimal Unary Encoding (OUE) mechanism. While OUE is a prominent method for item mining, it results in a high communication load, particularly when the perturbation domain is large. In future work, we plan to explore and develop mechanisms that can reduce this communication load for multi-class item mining. 
%\section*{Acknowledgment}
%This work was supported by the National Natural Science Foundation of China (Grant No: 62372122, 92270123, 62072390 and 12371522), and the Research Grants Council, Hong Kong SAR, China (Grant No:  15225921, 15209922 and 15208923).
\vspace{-0.5em}
\section*{Acknowledgment}
This work was supported by the National Natural Science Foundation of China (Grant No: 92270123 and 62372122), and the Research Grants Council, Hong Kong SAR, China (Grant No:  15225921 and 15208923). Dr. Qi Wang was supported by the National Natural Science Foundation of China (Grant No. 62250710682).

%\bibliographystyle{IEEEtran}
%\bibliography{ref}
\end{document}